\documentclass[aps,prd,superscriptaddress,onecolumn,
preprintnumbers,
nofootinbib,showpacs]{revtex4-2}
\usepackage{graphicx} 
\usepackage{appendix}
\usepackage{epsfig}
\usepackage{amssymb}
\usepackage{amsmath}
\usepackage{amsfonts}
\usepackage{anysize}
\usepackage{verbatim}
\usepackage{float}
\usepackage{cancel} 
\usepackage{framed}
\usepackage{mathrsfs}
\usepackage{amsbsy}
\usepackage{setspace}
\usepackage{color}
\usepackage{ulem}
\newcommand{\bea}{\begin{eqnarray}}
\newcommand{\eea}{\end{eqnarray}}
\newcommand{\be}{\begin{equation}}
\newcommand{\ee}{\end{equation}}
\newcommand{\nn}{\nonumber}
\begin{document}

\title{The real-time QCD static potential at high temperature}

\author{Margaret E. Carrington}
\affiliation{Department of Physics, Brandon University,
Brandon, Manitoba R7A 6A9, Canada}
\affiliation{Winnipeg Institute for Theoretical Physics, Winnipeg, Manitoba, Canada}
\author{Cristina Manuel}
\affiliation{Instituto de Ciencias del Espacio (ICE, CSIC) \\
C. Can Magrans s.n., 08193 Cerdanyola del Vall\`es, Catalonia, Spain}
\affiliation{Institut d'Estudis Espacials de Catalunya (IEEC) \\
08860 Castelldefels (Barcelona), Catalonia, Spain
}
\author{Joan Soto} 
\affiliation{Departament de F\'\i sica Qu\`antica i Astrof\'\i sica and Institut de Ci\`encies del Cosmos, 
Universitat de Barcelona, Mart\'\i $\;$ i Franqu\`es 1, 08028 Barcelona, Catalonia, Spain}
\affiliation{Institut d'Estudis Espacials de Catalunya (IEEC) \\
08860 Castelldefels (Barcelona), Catalonia, Spain
}
\date{\today}

\begin{abstract}

We develop a procedure to analytically calculate higher-order contributions to the high-temperature real-time static potential in QCD. It is based on the introduction of a semi-hard external scale, which lies between the hard scale (the temperature) and the soft scale (the screening mass), and the method of integration by regions.
We calculate the leading and next-to-leading corrections 
in the  region where bound states transit from narrow resonances to wide ones. 
The calculation involves both loop diagrams calculated in the Hard Thermal Loop (HTL) effective theory and power corrections to the HTL Lagrangian calculated in QCD. We also calculate the thermal corrections to the heavy quarkonium spectrum, and estimate the dissociation temperatures. We compare our results with recent lattice data and 
discuss 
their usefulness 
to guide lattice inputs in inverse problems.

\end{abstract}

\maketitle

\large

\maketitle

\section{Introduction}

The finite-temperature QCD static potential is a basic ingredient to understand the behavior of heavy quarkonium in a thermal medium. It was early realized that the Coulomb-like behavior at short distances would be screened in the medium in an analogous way to that of the Coulomb potential for charged particles. As a consequence, fewer and fewer bound states would be supported as the temperature increases, which leads to the conclusion that the production of charmonium (and eventually bottomonium) would be suppressed in heavy ion collision (HIC) experiments   \cite{Matsui:1986dk}. It took some time to realize that the real-time finite-temperature static potential develops an imaginary part \cite{Laine:2006ns}. The imaginary part becomes parametrically larger than the real part before screening becomes sizable \cite{Escobedo:2008sy}. This implies that the disappearance of bound states occurs because they become wide resonances rather than because they are no longer  supported by the Yukawa-like potential. The suppression of heavy quark bound state production  is indeed observed in current HIC experiments, clearly in the bottomonium ($\Upsilon$) family \cite{ALICE:2020wwx,CMS:2018zza} and indirectly for charmonium  \cite{CMS:2017uuv} where recombination effects must be taken into account. Experimental results show that the peaks corresponding to bottomonium states become wider and fuse with the background as the multiplicity (temperature) increases, rather than sequentially disappearing with roughly the same width. This pattern is consistent with the idea that bound states disappear because they decay, and not because the potential is screened to the point that it becomes too shallow to support them.

The calculation presented in \cite{Laine:2006ns} is a leading-order (LO) calculation in Hard Thermal Loop (HTL) effective theory. This implicitly assumes that the typical momentum of the gluons exchanged between the heavy quarks is smaller than the temperature (the hard scale). 
The effect of higher order contributions on this picture was an important open question that has recently been addressed in ref. \cite{Carrington:2024ize}.
We present here details of this beyond leading order (BLO) calculation. It is based on the fact that a new scale exists ($p\sim (T m_D^2)^{1/3}$ where $T$ is the temperature and $m_D\sim g T$ is the screening mass) for which the imaginary part of the momentum space potential is of the same order as the real part at LO. This scale fulfills the condition $m_D \ll p \ll T$ if the QCD coupling constant $g$ is small enough.  For momenta of the order of this new scale, which we call semi-hard, one can exploit the inequalties that order the three scales. We use the method of integration by regions \cite{Beneke:1997zp,Smirnov:2012gma} to simplify the calculations.  
We first discuss the corrections to the momentum space potential and provide expressions for these corrections in two approximations, the fully expanded one, which consistently respects the hierarchy discussed above, and the damped approximation, which partially resums higher order contributions.
We then discuss the coordinate space potential for distances corresponding to the semi-hard scale. We show that the soft momentum contributions have a universal form. We calculate the semi-hard momentum contributions by Fourier transforming the momentum space potential calculated previously. Expressions for both the fully expanded and the damped approximations are provided. 
Next, as a first application, the fully expanded expressions are used to analytically calculate  corrections to the thermal energy shifts and thermal decay rates, by treating all corrections to the QCD Coulomb potential perturbatively. We then push our formulas beyond their strict applicability range and compare our results with  recent lattice data.
The reasonable description we obtain encourages us to propose that our results could serve as inputs for Bayesian methods \cite{Jarrell:1996rrw,Asakawa:2000tr, Rothkopf:2011ef, Burnier:2013fca, Burnier:2013nla}, which are needed to obtain the imaginary part of the real-time potential from lattice QCD \cite{Rothkopf:2011db,Burnier:2014ssa, Burnier:2015tda, Burnier:2016mxc, Lehmann:2020fjt, Boguslavski:2021zga, Bala:2021fkm,Dong:2022mbo,Bazavov:2023dci}. Since the LO HTL potential is often used as inputs, our results provide a potentially significant improvement.

We organize the paper as follows. In Section \ref{sec-lo} we describe the formalism and discuss the leading-order results of ref. \cite{Laine:2006ns}. In Section \ref{sec-nlo} we explain our approach. In Sections \ref{sec-pow-tony} and \ref{sec-ladd} we present the calculations of the self-energy diagrams, and the ladder and vertex diagrams respectively (see fig.~\ref{fig-diag}). In Section \ref{sec-count} we discuss our power counting. In Section \ref{sec-coord} we calculate the coordinate space potential and discuss the universal form of the soft contributions. In Section \ref{sec-pert} we present an analytic calculation of the thermal energy shifts and thermal decay widths. In Section \ref{sec-lattice} we compare our results to lattice calculations, and estimate the dissociation temperature. Section \ref{sec-concl} is devoted to the conclusions.

\section{The static potential for $p \ll T$ at LO}
\label{sec-lo}
In this section we give some details of the leading order calculation that will be useful to understand the higher order computations presented in the following sections. 

We work in the closed-time-path (CTP) formalism of thermal field theory (see the reviews \cite{Zhou-CTP,Rothkopf:2019ipj,Ghiglieri:2020dpq}) and we use an approach based on the HTL effective theory \cite{Braaten:1991gm}. 
The static quark and antiquark are (unthermalised) probe particles that couple to gluons from the time-ordered branch of the CTP contour. The vertices in the anti-time-ordered branch are only needed for the calculation of the gluon self-energy (see section \ref{sec-pow-tony}). We use  dimensional regularization throughout. Additional information about the definitions and notational conventions we use are given in appendix \ref{appendix-notation}. 

\begin{figure}[H]
\begin{centering}
\includegraphics[scale=0.80]{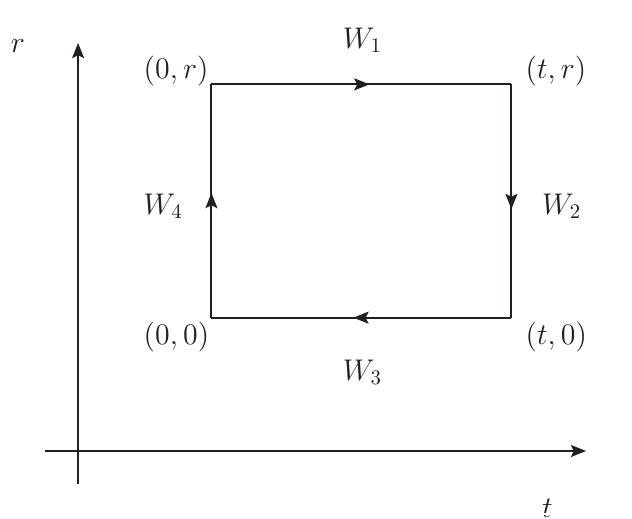} 
\caption{The Wilson loop used in eq.~(\ref{bigC}).\label{fig-Wloop}}
\end{centering}
\end{figure}

The static potential can be obtained from the Wilson loop  
\bea
W = {\cal P} \;{\rm Exp}\left(i g \oint dx^\mu A_\mu(x)\right) 
\label{W-def}
\eea
where the symbol ${\cal P}$ indicates path ordering along the rectangular path shown in fig.~\ref{fig-Wloop}. 
The potential is defined as
\bea
&& V(\vec r) = \lim_{t\to\infty} v(t,\vec r)  \nn \\
&& v(t,\vec r) = \frac{i}{t} \ln[C(t,\vec r)] \nn \\
&& C(t,\vec r) =  \frac{1}{N_c} \langle W_1 W_2 W_3 W_4 \rangle \label{bigC}
\eea
where we have used $W=W_1W_2W_3W_4$ and the notation $W_i$ refers to the Wilson line corresponding to each of the four branches of the loop as shown in fig.~\ref{fig-Wloop}. 
We calculate the potential perturbatively by expanding in $g$ to obtain
\bea
C_0(t,\vec r) + g^2 C_1(t,\vec r) + g^4 C_2(t,\vec r) + \dots=  {\rm Exp}\left[-it\left(g^2 v_1(t,\vec r) + g^4 v_2(t,\vec r) + \dots\right)\right]\,.
\label{C-exp}
\eea 
Matching orders in $g$ gives
\bea
 V_1(\vec r) &&= \lim_{t\to \infty} v_1(t,\vec r) = \lim_{t\to \infty} \left(\frac{i C_1(t,\vec r)}{t} \right) \  \label{Vlo} \\
 V_2(\vec r) &&= \lim_{t\to \infty} v_2(t,\vec r)    = \lim_{t\to \infty} \left(\frac{i C_2(t,\vec r)}{t} - \frac{i C^2_1(t,\vec r)}{2t}\right)\,. \label{Vorders}\label{Csplit}\label{Vnlo}
\eea
When we take the limit $t\to\infty$ the lines along the right and the left sides can be set to one. This means that the only component of the vector potential that contributes in eq.~(\ref{W-def}) is the zeroth component. 

The static potential is gauge invariant and therefore calculations can be done in any gauge. The fact that only the zeroth component of the gauge potential is relevant means that Coulomb gauge is a natural choice. We use strict Coulomb gauge and for some parts of our calculation we will verify our result by repeating the computation in Feynman gauge. 

The gluon propagator is defined from the equation
$\langle A_0(x)A_0(y)\rangle = -i  G(x-y)$
where the subscript $00$ on the propagator is suppressed throughout this paper. 
In momentum space the time-ordered propagator is obtained from the retarded and advanced propagators as 
\bea
G(p_0,\vec p) &=& G^{\rm ret}(p_0,\vec p) + n(p_0) (G^{\rm ret}(p_0,\vec p) - G^{\rm adv}(p_0,\vec p)) \label{timeordered} \\
& =&
\frac{1}{2}\left(G^{\rm ret}(p_0,\vec p) + G^{\rm adv}(p_0,\vec p) + (1+2n(p_0))\left[G^{\rm ret}(p_0,\vec p) - G^{\rm adv}(p_0,\vec p)\right]\right) \nn
\eea
where $n(p_0)$ is a Bose-Einstein distribution.
The HTL propagator in Coulomb gauge is given in appendix \ref{appendix-notation}. 
The static limit of the time-ordered component is
\bea
&& G(0,\vec p) = -\frac{1}{p^2+m_D^2} + \frac{i\pi T m_D^2}{p(p^2+m_D^2)^2}\,.
\label{static-htl}
\eea

To obtain the potential at leading order in the expansion of the Wilson lines in $g$, which we call $V_1(\vec r)$, we need to calculate $C_1(t,\vec r)$ (see eqs.~(\ref{bigC}, \ref{C-exp}, \ref{Vorders})). There are two contributions, one from the contraction of a gauge potential in the expanded link operator $W_1$ with a gauge potential from the expanded link operator $W_3$, and one from a contraction of two gauge potentials from the same line ($W_1$ or $W_3$). These two contributions are called respectively $C_{1a}(t,\vec r)$ and  $C_{1b}(t)$. 
The first type of contraction gives   
\bea
C_{1a}(t,r) &=& \frac{(ig)^2}{N_c} \left\langle \int_0^t dx_0 A_0(x_0,\vec r) \;\int^0_t dy_0 A_0(y_0,\vec 0)\right\rangle \nn \\
&=& -i g^2 C_F  \int_0^t dx_0 \int_0^t dy_0 \int \frac{d^4p}{(2\pi)^4} 
e^{-i(p_0(x_0-y_0)-\vec p\cdot\vec r)}G
(p_0,\vec p)\,. \label{oneC1a}
\eea
The second type of contraction, including a factor of two to account for two potentials from either $W_1$ or $W_3$, is
\bea 
C_{1b}(t) = 2 i g^2 C_F \int_0^t dx_0 \int_0^{x_0} dy_0 \int \frac{d^4p}{(2\pi)^4} e^{-i p_0(x_0-y_0)} G(p_0,\vec p) \,.\label{oneC1b}
\eea 
The potential $V_1$ is then obtained from the limit
\bea
V_{1}(r) = \lim_{t\to \infty}\frac{i}{t}(C_{1a}(t,r)+C_{1b}(t))\,.
\label{C1aC1b}
\eea
Using the results in appendix \ref{appendix-integrals} we easily find 
\bea
&& \lim_{t\to\infty}\frac{i}{t} C_{1a}(t,r)  = g^2 C_F  \int \frac{d^3p}{(2\pi)^3}e^{i\vec p \cdot \vec r} G(0,\vec p)   \label{C1a-check}\\
&& \lim_{t\to\infty}\frac{i}{t} C_{1b}(t)  = - g^2 C_F  \int \frac{d^3p}{(2\pi)^3} G(0,\vec p) \,.\label{C1b-check}
\eea
Equation (\ref{C1b-check}) gives a constant contribution to the coordinate space potential.  This term is $ -\alpha C_F m_D$ and has the correct form so that the real part of the potential gives the Coulomb potential in the limit $r\to 0$. For $p$ soft and semi-hard $1/r$ the leading contribution of (\ref{C1a-check}) cancels (\ref{C1b-check}).  In our calculations we consistently drop all constant contributions to the potential. These terms, together with the contributions from the soft momentum scale $p\sim m_D$ that we will not calculate, will be obtained by fitting to lattice data (see sec.~\ref{sec-fit} for a detailed discussion). 
The leading order coordinate space potential is obtained by replacing the propagator in  (\ref{C1a-check}) with the HTL expression (\ref{static-htl}) which is calculated using the assumption $p \ll T$.  This gives \cite{Laine:2006ns}
\bea
V_{1{\rm LO}}(r) 
= \int \frac{d^3p}{(2\pi)^3}e^{i\vec p \cdot \vec r}\tilde  V_{1{\rm LO}}(p)
= g^2 C_F  \int \frac{d^3p}{(2\pi)^3}e^{i\vec p \cdot \vec r} \left(
-\frac{1}{p^2+m_D^2} + \frac{i\pi T m_D^2}{p(p^2+m_D^2)^2}
\right)\,.\label{v1lo-p}
\eea

\section{Contributions beyond leading order}
\label{sec-nlo}

To obtain an expression for the potential beyond leading order we should expand the Wilson lines $W_1$ and $W_3$ in (\ref{bigC}) to higher order in the coupling constant. In addition, at any order in the expansion of the Wilson lines, we must decide if and how to dress the propagators and vertices in the resulting momentum integrals (i.e. if we should use bare $n$-point functions, or HTL corrected ones, or something that includes higher order corrections). This decision depends on the momentum scale that we choose to focus on, which in turn will determine the range of $r$ where our coordinate space potential will be most reliable. 
To understand this we start with the observation that the Fourier transform of the real part of the leading order potential in (\ref{v1lo-p}) gives the familiar Yukawa form
\bea
{\rm Re}[V_{1{\rm LO}} (r)] =   -  \frac{ g^2 C_F  }{4\pi r} e^{-m_D  r }\,.
\eea
If we expand the Yukawa potential in $r$ the leading term is the Coulomb potential $V(r) = - g^2 C_F/(4\pi r)$, which is the result we would have obtained if we had expanded the propagator (\ref{static-htl}) in $m_D/p$ before doing the momentum integral. At large $r$ however the Yukawa potential does not agree well with the Coulomb potential, because the small momentum region of the integrand is not accurately represented by the expanded propagator. We also note that the exponential dependence of the Yukawa potential on $g$ is inherently non-perturbative, which shows that we should not always think in terms of a simple expansion in powers of $g$. 

To decide what momentum scale is important for our purposes we return to eq.~(\ref{v1lo-p}) which shows that if $p\sim g^a T$ then
\bea
{\rm Re}[\tilde V_{1LO}(p)] \sim \frac{g^{2-2a}}{T^2}~~\text{and}~~{\rm Im}[\tilde V_{1LO}(p)] \sim \frac{g^{4-5a}}{T^2}\,.
\eea
For a narrow resonance to exist, the imaginary part of the static potential must be smaller than the real part, which implies $0<a<2/3$. 
We will describe the scale $g^a T$ with $0<a<2/3$ as semi-hard. 
The value $a=2/3$ gives the momentum scale for which the real and imaginary parts of the leading order  momentum space potential are of the same order and is parametrically the scale at which we expect quarkonium to dissociate. 
The coordinate space potential that we obtain upon Fourier transforming the momentum space potential will not be valid for $r m_D\gg 1$, since large $r$ corresponds to momenta softer than the semi-hard scale we have chosen, or  $r T\ll 1$, 
since small $r$ corresponds to momenta larger than the temperature. 
We note that since $T \gg p \gg m_D$ we can simultaneously satisfy the conditions that perturbation theory is valid and that narrow resonances exist. This hierarchy of scales could be exploited using non-relativistic effective field theories \cite{Caswell:1985ui,Pineda:1997bj,Brambilla:2008cx} (see \cite{Brambilla:2004jw,Pineda:2011dg} for reviews), but we  use instead the technique of integration by regions \cite{Beneke:1997zp,Smirnov:2012gma}, which is better suited for our purposes. 

\begin{figure}[H]
\begin{center}
\includegraphics[width=0.75\textwidth]{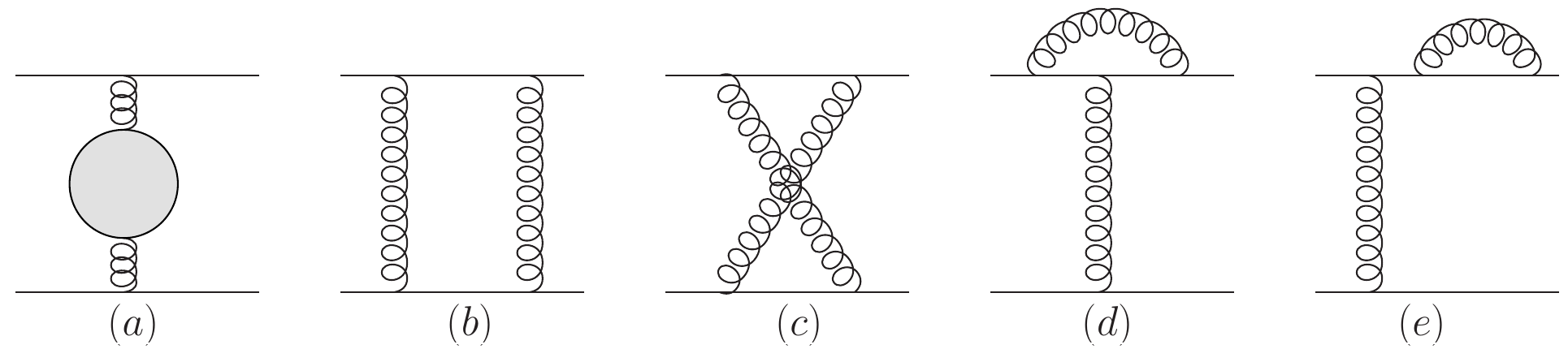}
\caption{One-loop contributions to the static potential in  Coulomb gauge. All gluon lines correspond to longitudinal gluons. The iteration of the LO potential must be subtracted. 
\label{fig-diag}}
\end{center}
\end{figure}

The relevant diagrams are shown in fig.~\ref{fig-diag} where all the fermion lines do not depend on the spatial momenta (the static limit). 
The first diagram has the same structure as the one that produced $V_1$ and the next four come from higher order terms in the expansion of the Wilson lines. In all of these diagrams there is one (4-dimensional) momentum variable that is integrated over (which we call $k$) and the momentum transfer is $p=(p_0,\vec p)$, as in the calculation of $V_1$ in sec.~\ref{sec-lo}. The energy transfer $p_0$ is eventually taken to zero to get the momentum space potential. Fourier transforming the momentum space potential gives the coordinate space potential. 

The contribution from the first diagram in fig.~\ref{fig-diag} is called $V_{1{\rm NLO}}$. The blob in the middle of the diagram is the gluon self-energy calculated beyond the leading order HTL approximation. 
There are two relevant contributions. One comes from the power correction to the HTL self-energy ($k\sim T$) and one from the one-loop self-energy diagram with the loop momentum  semi-hard $k\sim p$. In this calculation, the fact that the semi-hard scale dominates the soft one ($m_D$) produces enormous simplifications. Both HTL vertices and HTL propagators reduce to the corresponding bare ones up to corrections of order $m_D^2/p^2$. There are no contributions 
from quark loops for $k\sim p$ because at energies smaller than $T$ they are Pauli-blocked and hence suppressed by $p^2/T^2$ with respect to the Bose-enhanced gluonic contributions. Also possible contributions from tadpole diagrams vanish. 

We refer to the last four graphs in fig.~\ref{fig-diag} as the ladder ((b) and (c)) and vertex ((d) and (e)) graphs.
We will calculate them only in Coulomb gauge. 
The ladder and vertex graphs are zero at zero temperature in Coulomb gauge because they involve only the bare longitudinal gluon propagator which is energy independent. 
At finite temperature they give a nonvanishing contribution that has not previously been considered. The leading contribution arises when the internal momentum $k\sim m_D$ and HTL propagators are used, because the HTL longitudinal gluon
propagator is energy dependent. 
We note that there is no diagram analogous to fig.~\ref{fig-diag}d with a three-gluon vertex.
This diagram is missing because there is no off-diagonal propagator connecting the longitudinal ($A_0$) and
transverse ($A_i$) fields in Coulomb gauge and no vertex with three $A_0$. 

The calculation of these diagrams is explained in detail in sections \ref{sec-pow-tony} and \ref{sec-ladd}. Calculations are done by exploiting the hierarchy $m_D\ll p \ll T$ and organized by powers of the coupling using $m_D\sim gT$ and $p\sim g^a T$ with $0<a<2/3$. 
To decide what terms to keep we must find the order of the largest terms that are not included in our calculations. Using power counting we can find the orders of the 2-loop contributions to the potential and the corrections we would find to the 1-loop diagrams we have calculated if we had used HTL corrected vertices. The orders of these corrections are
\bea
&& \text{2 loop contributions:~~} \text{Re}[V] \sim g^{4-2a} \text{~~and~~} \text{Im}[V]\sim g^{4-2a}  \nn \\
&& \text{HTL vertex contributions:~~} \text{Re}[V] \sim g^{6-5a} \text{~~and~~} \text{Im}[V]\sim g^{6-6a} \,. \label{dropped-this}
\eea
In appendix \ref{sec-orders} we explain how to obtain these expressions. 
We will take into account all contributions to the potential larger than those shown in eq.~(\ref{dropped-this}). 

\section{Corrections to the gluon self-energy}
\label{sec-pow-tony}

In this section we provide the corrections to the HTL gluon self-energy which are indicated by the grey blob in the first graph in fig.~\ref{fig-diag}. There are two kinds of contributions: one-loop diagrams with loop momenta $k \sim p$ and power corrections to the lowest order HTL gluon self-energy. We give some details on how these contributions are calculated in the following two sections. 

\subsection{ HTL corrections for $m_D \ll p \ll T$}
 \label{sec-tony}
 
 The leading correction to the HTL gluon self-energy comes from semi-hard loop momenta because of a Bose-Einstein enhancement of the gluon contribution. If the external momentum were soft we would need to use HTL propagators and vertices. Since we consider semihard external momenta the calculation is much simpler. For external momenta that are semi-hard  we can expand both the HTL propagators and vertices in powers of $m^2_D/p^2$ and to the order we work we will need only the first term (bare lines and vertices). In the gluon distribution function we can make the approximation $n (p_0) \sim T/p_0$
which is equivalent to considering the classical field theory limit in the gauge sector.
We describe below the calculation in Coulomb gauge. In appendix \ref{appendix-feyn} we show that the same result is obtained in Feynman gauge.\footnote{In Feynman gauge there are additional contributions from the vertex diagrams that must be included.} The tadpole is zero in DR so we only need the contribution from the gluon bubble. The Feynman rules in Coulomb gauge are given in appendix \ref{appendix-notation}. After contracting all indices we can shift variables so that all terms in the integrand contain a delta function with argument equal to the square of the loop four-momentum. We use this delta function to do the frequency integral. The result is
\bea
[\Pi_{00}(p_0,\vec p)]_{\rm semi-hard} &=& -\frac{1}{2}g^2N_c T  \int \frac{d^dk}{(2\pi)^d}
\bigg[\frac{8 p^2 \left(1-x^2\right)}{|\vec k+\vec p|^2 k^2} 
+ \frac{1}{k^2} 
\left(d-1 - \frac{p^2 \left(1-x^2\right)}{|\vec k+\vec p|^2}\right)
\label{htl1}\\
&& 
\left(
\frac{\left(p_0-2 k\right){}^2}{P^2 -2 k \left(p x+p_0\right) - 2 i \eta  \left(k-p_0\right)}
+\frac{\left(2 k+p_0\right){}^2}{P^2+2 k \left(p_0-p x\right)+ 2 i \eta  \left(k+p_0\right)}
\right)
\bigg]\,\nn
\eea
where we use the notation $P^2=p_0^2-p^2$ and $\eta$ denotes a positive real infinitesimal.
We expand in $p_0$ because we only need the polarization tensor in the limit $p_0\to 0$. 
Since the angular integrals and the integral over the magnitude of $\vec k$ are separately finite we set $d=3$ in the integrand and write (using $d=3+2\epsilon$) 
\bea
\int \frac{d^dk}{(2\pi)^d} \to \frac{1}{(2\pi)^2} \int_{-1}^1 dx \, \int dk \; k^{2+2\epsilon}\,.
\eea
Then we rearrange the integrand in a way that makes it easier to do the angular integrals. 
First, for each term, we remove factors of $x$ in the numerator by writing them in terms of the corresponding denominator. Then we partial fraction terms with denominators that have products of different factors that depend on $x$. Keeping terms of order $p_0^1$ the square bracket in (\ref{htl1}) can be written in the form 
\bea
[~~] &\to& \sum_{j\in\{-1,1\}}\left\{
\frac{k( 4 k^2 p^2 - p^4- 8 k^4 ) + jp_0(8 k^4 -p^4)}{k^5 \left(p^2 +2 k p x + j i  \eta \right)}
+ \frac{2 j p_0 \left(8 k^4-4 k^2 p^2+p^4\right)}{k^3 \left(p^2 +2 k p x + j i \eta \right)^2}\right\}
\,.
\label{sq1}
\eea
Using the expression for the integrand in (\ref{sq1}) it is easy to do the $x$ integrals. The result is
\bea
[\Pi_{00}(p_0\to 0,\vec p)]_{\rm semi-hard} &=& -\frac{1}{2}g^2 N_c T  \frac{1}{(2\pi)^2} \frac{1}{p} \int dk k^{2+2\epsilon}\nn \\
&& \bigg[ ~~\frac{p_0}{k^2} 
   \left\{\frac{p^4}{2 k^4}-4\right\}  \sum_{jm} m \ln(1+2jk/p+i m \eta) \nn \\
&& - \frac{1}{k}  \left\{\frac{p^4}{2 k^4} - \frac{2 p^2}{k^2} + 4 \right\}  \sum_{jm}  j\ln(1+2jk/p+i m \eta) \nn\\
&& - \frac{2p_0}{k}\left\{\frac{p^2}{k^2} - 4 + \frac{8k^2}{p^2} \right\}\sum_{jm}\frac{mj}{p+2kj+im\eta}~~ \bigg]+ {\cal O}(p_0^2)
\eea   
where the indices $j$ and $m$ are summed over $\{-1,1\}$.
The remaining $k$~integral can be easily done and gives 
\bea
[\Pi_{00} (p_0\to 0, p)]_{\rm semi-hard} = -\frac{g^2 N_c T }{4} \left( p  + \frac{ 7 i  }{3\pi} p_0 \right) + {\cal O}(p_0^2)\,. \label{tony-result}
\eea
The real part of this result was calculated previously in \cite{Rebhan:1993az} and the imaginary part in \cite{Shi:2015tmz,Zhu:2015edf}.
When we substitute (\ref{tony-result}) into the time-ordered propagator and expand in $p/T$ we get corrections to the LO HTL propagator (\ref{static-htl}) and the momentum space potential of the form
\bea
&& \tilde V_{1{\rm HTL}}(p) = g^2 C_F \, G_{\rm nlo}(0,p) \nn \\[2mm]
&& G_{\rm nlo}(0,p)  = -\frac{g^2 N_c p\, T}{4(p^2+m_D^2)^2}   -\frac{i g^2 N_c \,\frac{7}{3}\,T^2}{2\pi(p^2+m_D^2)^2} \,.\label{prop-HTL}
\eea

\subsection{Power corrections to the HTL gluon self-energy}
\label{sec-pow}

Subleading corrections to the HTL gluon self-energy arise as power corrections (POW). These corrections are obtained from one-loop diagrams by
assuming that the external semi-hard momentum is smaller than either the hard loop momenta or the temperature and expanding the one-loop integrand. 
We give below the result for the POW associated with the longitudinal part of the gluon polarization tensor in Coulomb gauge.

The POW from the quark contribution can be obtained from the QED expression,  after simple generalizations, and is obviously gauge independent.
The quark contribution is 
\cite{Manuel:2016wqs,Carignano:2017ovz}
\begin{equation}
\Pi^{\rm POW}_q (p_0, p)= - \frac{g^2 N_F}{ 24 \pi^2 }\left[ \frac{p^2}{ \epsilon} +  2 p^2
 \left( \ln \left[\frac{\sqrt{\pi}Te^{- \gamma/2}}{2 \mu}\right]  -1 \right) +
\left(3 p^2-p_0^2\right)\left(1- \frac{p_{0}}{2 p} 
 \,{\rm ln}\left({\frac{p_0+ p+i\eta}{p_0-{ p}+i\eta}}\right) \right) \right]  \label{final-L}
\end{equation}
where $\mu$ is the renormalization scale and $\gamma$ is Euler's constant.

The contribution from the gauge degrees of freedom can be calculated in a straightforward way by expanding the one-loop integrand. 
The leading terms give the HTL contribution. The sub-leading correction is the POW and reads
\begin{equation}
\Pi^{\rm  POW} _g(p_0, p) =   \frac{g^2 N_c}{48 \pi^2} p^2 \left[ 11 \left( \frac 1\epsilon -1 - \ln\left[{\frac{\mu^2 e^\gamma}{T^2 4 \pi}}\right] \right) -  \frac{14p_0^2}{p^2}  
+ \frac {p_0 ( 7 p_0^2-6 p^2)}{p^3} \ln\left( { \frac{p_0+p+i\eta}{p_0-p+i\eta} }\right) \right]\,.
\label{gluon-pow}
\end{equation}
From eqs.~(\ref{final-L}, \ref{gluon-pow}) it is easy to see that the leading contribution to the POW is $\sim g^2p^2$ for the real part and $\sim g^2 p_0 p$ for the imaginary part. The corresponding contributions to the static potential will be 
\bea
&& {\rm Real:~~} \frac{g^2(g^2p^2)}{p^4} \sim g^{4-2a} \nn \\
&& {\rm Imag:~~}\frac{g^2(g^2 p_0 p)}{p^4} \frac{T}{p_0}\sim g^{4-3a} \,.\nn
\eea
Comparing with (\ref{dropped-this}) we see that the real part of the POW does not contribute at the order we are working at. 
After adding the quark and gluon contributions we get for the imaginary part of the POW
in Coulomb gauge
\begin{equation}
\label{IM-POW}
{\rm Im}\, \Pi^{\rm  POW} (p_0,p) \approx  \frac{ g^2 }{ 8 \pi} \left( N_c - \frac{N_F}{2} \right) p_0 p  \ .
\end{equation}
Inserting this result into the time ordered propagator (\ref{timeordered}) and expanding in $g$ we find the corresponding contribution to the momentum space potential 
\bea
&& \tilde V_{1{\rm POW}}(p) = \frac{i g^4 p T C_F \left(N_c-N_f/2\right)}{4 \pi  \left(m_D^2+p^2\right){}^2}\,.
\eea
 
\section{Ladder and vertex graphs}
\label{sec-ladd}

As explained in sec.~\ref{sec-nlo}, the leading contribution from these graphs arises when the internal momentum $k\sim m_D$ and HTL propagators are used. We also must include the iteration of the leading order contribution. 
From eq.~(\ref{Csplit}) we find 
\bea
V_2 = \lim_{t\to\infty} \left[
\left\{\frac{i C_{2b}(t,\vec r)}{t} - \frac{i C^2_{1a}(t,\vec r)}{2t}\right\}
+\frac{i C_{2c}(t,\vec r)}{t} 
+\frac{i C_{2d}(t,\vec r)}{t} 
+\left\{\frac{i C_{2e}(t,\vec r)}{t} - \frac{i}{t} C_{1a}(t,\vec r)C_{1b}(t) \right\}\right]\,.\nn \\
\label{curly}
\eea
Note that an overall constant from a term in the square bracket $\sim C^2_{1b}(t)$ has been dropped. We have also dropped constant terms in the potential from diagrams not  included in fig.~\ref{fig-diag} that correspond to static quark self-energy contributions. 
In sec.~\ref{sec-fit} we discuss how we determine these constant contributions in our calculation. 
The notation $C_{2b}$, $C_{2c}$, $C_{2d}$ and $C_{2e}$ indicates different contributions to $C_{2}$ that correspond to the second, third, fourth and fifth diagrams in fig.~\ref{fig-diag}. The curly brackets group together terms that are free from pinch singularities. 
The contributions $C_{2b}$ and $C_{2c}$ come from the second order terms from the expanded exponentials of $W_1$ and $W_3$ which give
\bea
[C_2(t)]_{2\times 2} = \frac{(ig)^4}{N_c} \xi_{abcd}  \int_0^t dx_0 \int_0^{x_0} dy_0 A_0^a(x_0,\vec r)A_0^b(y_0,\vec r) 
\int^0_t dz_0 \int_t^{z_0} dw_0 A_0^c(z_0,\vec 0)A_0^d(w_0,\vec 0) \,
\label{22}
\eea
where we have defined $\xi_{abcd} = {\rm Tr}(T^aT^bT^cT^d)$. 
The graphs in fig.~\ref{fig-diag}b and fig.~\ref{fig-diag}c come from different contractions of  the potentials in eq.~(\ref{22}). 
The contributions in fig.~\ref{fig-diag}d and fig.~\ref{fig-diag}e are obtained from taking the third order term from $W_1$ and first order from $W_3$ which gives
\bea
[C_2(t)]_{3\times 1}  = \frac{g^4\xi_{abcd}}{N_c} \int_0^t dx_0 \int_0^{x_0} du_0 \int_0^{u_0} dy_0  A_0^a(x_0,\vec r)A_0^b(u_0,\vec r) A_0^c(y_0,\vec r)
\int^0_t dz_0 A_0^d(z_0,\vec 0) \,.\label{31} 
\eea
Once again the graphs in figs.~\ref{fig-diag}d and \ref{fig-diag}e are obtained from different contractions of the potentials. In appendix \ref{appendix-pinch} we give full details of the calculation and show that the results are
\bea
&& V_2^{bc}(r)
 = -\frac{i g^4 N_c\,C_F}{2}
\int\frac{d^3p}{(2\pi)^3} e^{i \vec p \cdot \vec r} \int\frac{d^4k}{(2\pi)^4} 
\frac{ G(k_0,\vec k+\vec p) G(k_0,k) }{(k_0+i\eta)^2}
\, \label{ladd-cross-all} \\[2mm]
&& V_2^{de}(r)  
= ig^4 N_c \,C_F \int \frac{d^3p}{(2\pi)^3}e^{i\vec p \cdot \vec r}G(0,p) \int \frac{d^4k}{(2\pi)^4}\frac{G(k_0,k)}{(k_0+i\eta)^2}\,.
\label{vert-wave-all}
\eea
In these results we have included the appropriate symmetry factors which account for the vertex correction in fig.~\ref{fig-diag}d on the upper or lower line, and the self-energy correction in fig.~\ref{fig-diag}e to the right or left of the rung, and on the upper or lower line.

In eq.~(\ref{vert-wave-all}) the two momentum integrals are decoupled.  
To calculate the $k$-integral we rewrite it using eq.~(\ref{timeordered}) in the form 
\bea
&& \int \frac{d^4k}{(2\pi)^4} \frac{1}{(k_0+i\eta)^2}\;G(k_0,k) =   \int \frac{d^4k}{(2\pi)^4}
\frac{1}{(k_0+i\eta)^2}\left( \frac{1}{2} (r_k+a_k) +  
\frac{T}{(k_0+i\eta)}  (r_k- a_k)\right)\, \label{II-vert3}
\eea
where we use the shorthand notation $r_k \equiv G^{\rm ret}(k_0,\vec k)$ and $a_k\equiv G^{\rm adv}(k_0,\vec k)$. The term proportional to $T$ is obtained by expanding the distribution function (using $k\sim m_D \ll T$) and using that $(r_k-a_k)/k_0$ is analytic when $k_0\to 0$  which means we can write
\bea
 \frac{1}{k_0}(r_k-a_k) \to \frac{1}{k_0+i\eta}(r_k-a_k)\,.
\eea
The integrals can be done in a straightforward way using Cauchy's theorem and give
\bea
\int \frac{d^4k}{(2\pi)^4} \frac{1}{(k_0+i\eta)^2}\;G(k_0,k)  =  -\frac{1}{16\pi}  - \frac{i  T}{8 \pi  m_D}\left(1-\frac{3\pi}{16}\right) \,.
\label{Y-res}
\eea

Now we consider $V_2^{bc}(r)$ in eq.~(\ref{ladd-cross-all}). 
To do the integral we expand in $p\gg (k_0,k)$ but we keep factors of $m^2_D$ in the denominators, even though the Debye mass is parametrically smaller than $p$. 
Since we have assumed $p\gg m_D$ this procedure is not fully consistent
but it extends the region where we expect our coordinate space
potential to be valid, as explained in the first paragraph of sec.~\ref{sec-nlo}. 
From eqs. (\ref{ladd-cross-all}, \ref{vert-wave-all}) it is easy to see that the leading order term is
\bea
[V_2^{bc}(r)]_{lo} = -\frac{1}{2}V_2^{de}(r)\,.
\label{V2-dom}
\eea
Equations (\ref{static-htl}, \ref{vert-wave-all}, \ref{Y-res}, \ref{V2-dom}) give the dominant contribution to $V_2$ from the ladder graphs, which is pure real
\bea
\tilde V_{2}^{(0)}(p)
= -\frac{g^4 N_c C_F}{16\pi} \frac{1}{p^2+m_D^2}\left(\left(1-\frac{3\pi^2}{16}\right)\frac{T}{m_D}
\right)\,.
\label{vert-wave-2}
\eea
The next order correction to (\ref{V2-dom}) comes from the subleading term in the expansion in $p\gg (k_0,k)$ and is
\bea
\tilde V_{2}^{(1)}(p)
&& = \frac{g^4 T N_c C_F}{16}\frac{m_D}{\left(m_D^2+p^2\right){}^2} 
\left( \frac{1}{3 \pi  }\left(1-\frac{\pi ^2}{16}\right)
+ \frac{i  T}{p }\left(1-\frac{3 \pi ^2}{16}\right) \right) 
\,.\label{damp-kdone}
\eea
\section{Power counting}
\label{sec-count}

We collect below the results from sections \ref{sec-pow-tony} and \ref{sec-ladd}. 
\bea
&& {\rm Re}[\tilde V_2(p)] = {\rm Re} [\tilde V_2^{(0)}(p) + \tilde V_{1{\rm HTL}}(p) + \tilde V_2^{(1)}(p)]  \nn \\[4mm]
&& {\rm Re} [\tilde V_2^{(0)}(p)] = -\frac{g^4 N_c C_F}{16\pi} \frac{1}{p^2+m_D^2}\left(1-\frac{3\pi^2}{16}\right)\frac{T}{m_D}
\label{re20}\\ [4mm]
&& {\rm Re} [ \tilde V_{1{\rm HTL}}(p)] = \frac{g^4 p T N_c C_F}{4 \left(m_D^2+p^2\right)^2}\label{retony}\\[4mm]
&& {\rm Re} [\tilde V_2^{(1)}(p)] = \frac{g^4 T N_c C_F}{16}\frac{m_D}{3\pi\left(m_D^2+p^2\right)^2} 
\left(1-\frac{\pi ^2}{16}\right)\label{re21}\\[4mm]
&& i{\rm Im}[\tilde V_2(p)] = i{\rm Im} [\tilde V_{1{\rm HTL}}(p)]  +i{\rm Im} [\tilde V_{1{\rm POW}}(p)]+i{\rm Im} [\tilde V_2^{(1)}(p)] \nn \\[4mm]
&& i{\rm Im} [\tilde V_{1{\rm HTL}}(p)] 
=  -\frac{7 i g^4 T^2 N_c C_F}{6 \pi  \left(m_D^2+p^2\right){}^2}
\label{imtony}\\[4mm]
&& i{\rm Im} [\tilde V_{1{\rm POW}}(p)] = \frac{i g^4 p T C_F \left(2 N_c-N_f\right)}{8 \pi  \left(m_D^2+p^2\right){}^2}
\label{impow}\\[4mm]
&& i{\rm Im} [\tilde V_2^{(1)}(p)] = \frac{g^4 T N_c C_F}{16}\frac{i  T m_D}{p \left(m_D^2+p^2\right)^2} 
\left(1-\frac{3 \pi ^2}{16}\right) 
\,.\label{im21}
\eea
We must show that the smallest of these contributions is still larger than the biggest of the contributions that we did not calculate, which are given in eq.~(\ref{dropped-this}). When we make these comparisons we remember that we have assumed that the momentum scale is semi-hard $p\sim g^a T$ with $0<a<2/3$ since this is  the scale at which we expect quarkonium to dissociate.

For the imaginary part we consider separately the regions $0<a<1/2$ and $1/2<a<2/3$. 
For $a<1/2$ the smallest term we keep is Im$[\tilde V_{2}^{(1)}(p)]\sim g^{5-5a}\,T$ and the largest term not calculated is the 2-loop correction $\sim g^{4-2a}\,T$, which means that we require $a>1/3$. For $a>1/2$ the smallest term we keep is 
Im$[\tilde V_{1{\rm POW}}(p)]\sim g^{4-3a}\,T$ 
and the biggest terms not calculated are the HTL vertex corrections $\sim g^{6-6a}\,T$, which are always smaller for $0<a<2/3$. 
The smallest contribution to the real part of the potential that we have calculated is the subleading ladder term Re$[\tilde V_{2}^{(1)}(p)]$ in (\ref{re21}) which is $\sim g^{5-4a}\,T$. The biggest terms we have dropped are $\sim g^{4-2a}\,T$ and therefore we must require $a>1/2$. 

Combining these conditions we must require $1/2<a<2/3$. We also note that the first term in (\ref{Y-res}) has been omitted because the contributions it gives to the final result are smaller than the corrections we do not calculate (see eq.~(\ref{dropped-this})). Finally we comment that the leading order contribution to the real part of the potential from the ladder graphs Re$[\tilde V_{2}^{(0)}(p)]$ is parametrically larger than the contribution from Re$[\tilde V_{1{\rm HTL}}(p)]$ which was calculated previously \cite{Rebhan:1993az}. 

\section{The static potential in coordinate space for $1/r \gg m_D$}
\label{sec-coord}
 \subsection{$ m_D \ll p \ll T $ contributions}
 \label{sec-soft}

We Fourier transform the momentum space potential to obtain the potential in coordinate space. 
We define $\hat r = m_D r$ and use the notation
\bea
&& \hat I_j(\hat r) = \int_0^\infty d\hat p \, \frac{\sin \left(\hat{p} \hat{r}\right)}{\left(\hat{p}^2+1\right)^j}\label{meijerG}\\[2mm]
&& \hat {\cal I}_1(\hat r) = \frac{2 \hat I_1(\hat r)}{2\,\hat r \ln(\frac{e^{1-\gamma_E}}{\hat r})} \text{~~and~~} \hat {\cal I}_2(\hat r) = \frac{2\hat I_2(\hat r)}{\hat r } 
\eea
where the integrals $\hat {\cal I}_j(\hat r)$ are  constructed from $\hat I_j(\hat r)$  so that they are $\approx 1$  at small $\hat r$.

\begin{figure}[H]
\centering
\includegraphics[width=0.45\linewidth]{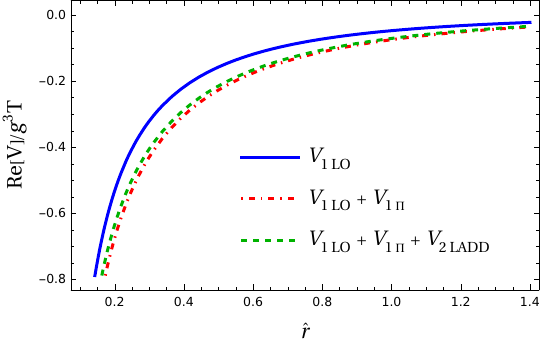}
\includegraphics[width=0.45\linewidth]{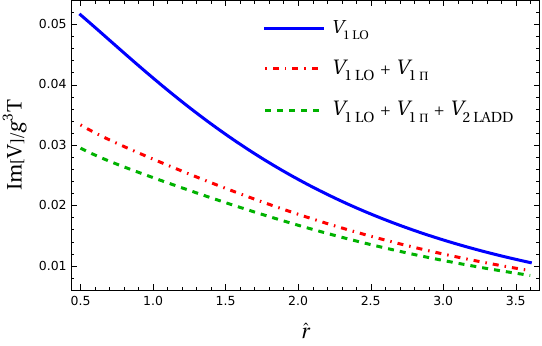}
\caption{The real and imaginary parts of the leading order potential and the BLO corrections using $V_{1\Pi} = V_{1{\rm HTL}}+ V_{1{\rm POW}}$ and 
$V_{2{\rm LADD}} = V_2^{(0)}+V_2^{(1)}$ and $g=1.8$. \label{plots-ana}}
\end{figure}

In coordinate space the results are:
\bea
&& {\rm Re}[V_{1{\rm LO}}] =   - g^3 \frac{C_F \hat{m}_D }{4\pi}  \frac{T\,e^{-\hat r} }{ \hat r} \, 
\label{mikko-real-2} \\
&& {\rm Re}[V_{1{\rm HTL}}] =  -g^4 N_c C_F  \frac{ T }{16 \pi ^2 }  
\left(2\ln\left(\frac{e^{1-\gamma_E}}{\hat r}\right)\hat{\cal I}_1(\hat r) - \hat{\cal I}_2(\hat r) \right) \\
&& {\rm Re}[V_2^{(0)}] = \frac{g^4 T N_c C_F}{64 \pi ^2} \bigg(
\frac{e^{-\hat{r}}}{\hat r} \bigg(\frac{3 \pi ^2}{16}-1\bigg)
\bigg)\,\\
&& {\rm Re}[V_2^{(1)}] = \frac{g^4 T N_c C_F}{384 \pi ^2} 
e^{-\hat{r}} \left(1-\frac{\pi ^2}{16}\right)\,\\
&& i {\rm Im}[V_{1{\rm LO}}]  
=  i g^2 C_F \frac{T}{4\pi}\, \hat {\cal I}_2(\hat r)
\label{mikko-imag-2} \\
&& i {\rm Im}[V_{1{\rm HTL}}]
 = - i g^3 \frac{C_F \,N_c\,}{16\pi^2  \hat{m}_D}\,\frac{7}{3} \; T e^{-\hat r} \label{extra-2}\\
&& i{\rm Im}[V_{1\rm POW}] = \frac{i g^4 T C_F \left(2 N_c-N_f\right)}{32 \pi ^3} \left(2
\ln\left(\frac{e^{1-\gamma_E}}{\hat r}\right)\hat{\cal I}_1(\hat r) 
- \hat{\cal I}_2(\hat r) \right) \\ \label{dampladder}
&& i{\rm Im}[V_2^{(0)}]  = 
- \frac{i g^3  T N_c C_F }{64 \pi ^2 
   \hat{m}_D} \left(\frac{3 \pi ^2}{16}-1\right) \hat {\cal I}_2(\hat r) \,\label{dampladder-t}
\eea
where we have defined $m_D=g T  \hat m_D$.
In fig.~\ref{plots-ana} we show the leading order potential and the BLO corrections we have calculated. In the figure we use $g=1.8$ which is obtained from ref.~\cite{Bazavov:2023dci} (see sec.~\ref{sec-rasmus}).

 \subsection{ $p \lesssim m_D$ contributions general parametrization}
 \label{sec-fit}
The contributions to the coordinate space potential from the soft region $(p \sim m_D)$
that we do not calculate have
a universal form at any order in $g$ for the region $m_D r\ll 1$. 
To see this we consider
\bea
V_2(r) = \int \frac{d^3p}{(2\pi)^3}e^{i\vec p\cdot\vec r}\tilde V_2(p) \nn
\eea
with $r\sim g^{-a}T^{-1}$ and $1/2<a<2/3$. 
 For soft momenta we can take advantage of the fact that $pr\sim m_D r \sim g^{1-a}<1$ to expand the exponential which gives
\bea
V_{2\,\rm soft}(r) &=& \int \frac{d^3p}{(2\pi)^3}\left(1+ i\vec p\cdot\vec r -\frac{1}{2} (\vec p\cdot\vec r)^2 +\cdots\right)\tilde V_2(p) 
\,.\label{V-soft}
\eea
The  terms in the round bracket with odd powers are zero by symmetry in an isotropic system. Note that an UV regulator may have to be introduced and the result would then be scale dependent (we use dimensional regularization so that all integrals are defined). 
If $p$ is soft then the biggest contribution to the imaginary part of $\tilde V_2(p)$ from the diagrams we have calculated is $\sim g^4 T^2/p^4$ where the $T^2$ is from two Bose enhancement factors and the $p^{-4}$ follows from dimensional analysis. The biggest contribution to the real part of $\tilde V_2(p)$ from the diagrams we have calculated is $\sim g^4 T/p^3$ because there can be only one Bose enhancement factor. 
This means that the contribution to the coordinate space potential from the soft momentum regime is of order 
\bea
&& {\rm Re}[V_2] \sim g^4 T (1+r^2 p^2+r^4 p^4+\dots) \sim T (g^4+g^{6-2a}+g^{8-4a} + \dots) \label{fit-real}\\[2mm]
&& {\rm Im}[V_2] \sim \frac{g^4 T^2}{p} (1+r^2 p^2+r^4 p^4+\dots) \sim T (g^3+g^{5-2a}+g^{7-4a} + \dots) \label{fit-imag}
\eea
where we have included a factor $p^3$ from the phase space of the momentum integral, and to get the expressions on the right side of both equations we used $p\sim m_D$ and $pr\sim g^{1-a}$. 

To determine which of these terms should be kept  we compare them with the terms we have included in our analytic calculation of the semi-hard contributions. We should only keep fitted terms that are greater than or equal to the smallest contributions we have included in our analytic result. 
Expanding 
eqs.~(\ref{re20} - \ref{im21}) in $m_D/p$ gives
\bea\label{expanded}
\tilde V_{2 \,{\rm exp}}(p)= - \frac{g^4C_F N_c T}{16\pi m_D p^2} &\bigg\{&
\left[1-\frac{3 \pi ^2}{16} +  \frac{4 \pi m_D}{p} + \frac{m_D^2}{p^2}\left(\frac{5\pi ^2}{24}-\frac{4}{3}\right)\right] \\[2mm]
&+& i\frac{\pi T m_D}{ p^2} \left[ 
\frac{56}{3 \pi} -\left(1-\frac{3 \pi ^2}{16}\right) \frac{m_D}{p} 
-\left(1-\frac{N_f}{2N_c}\right)\frac{4 p}{\pi T}\right]\bigg\} \,.\nonumber
\eea
To obtain this result we  keep terms of order $g^{5-4a}/T^2$ for the real part and $g^{5-5a}/T^2$ for the imaginary part 
(see the discussion at the end of sec.~\ref{sec-count}). 
We use $p\sim m_D$ and $r\sim g^{-a}T^{-1}$ and determine which fitted terms are greater than or equal to the terms that we have calculated. This shows we should keep the first term in (\ref{fit-real}) and 
the first two terms in (\ref{fit-imag}). The general form of these contributions in the coordinate space potential is therefore
\bea \label{soft}
V_{2\,{\rm soft}}=g^4 q_0 T 
+ i\left(g^3 i_0 T+ g^5 i_2 r^2 T^3\right)\,
\eea 
where $(q_0,i_0,i_2)$ are real constants, which may depend on the factorization scale. In sec.~\ref{sec-lattice} we will discuss how we obtain the numerical values of these constants by fitting to lattice data. 

\begin{figure}[H]
\centering
\includegraphics[width=0.45\linewidth]{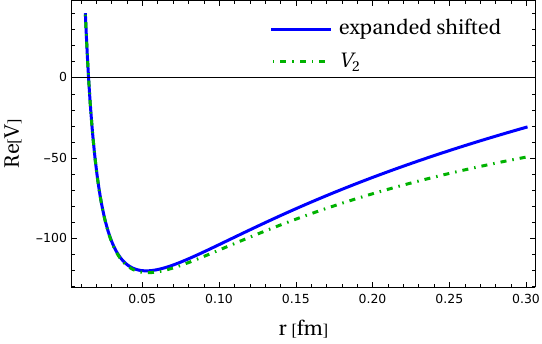}
\includegraphics[width=0.45\linewidth]{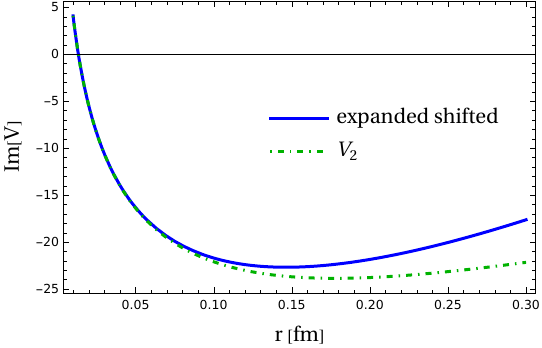}
\caption{The BLO potential and the expanded approximation for $g=1.8$ and $T=180$~MeV. The expanded approximation has been shifted to match the BLO potential at $r=0.01$~fm.  \label{plots-exp}}
\end{figure}

We also note that the soft contributions in (\ref{soft})  have the correct form to absorb the IR poles in the Fourier transform of the expanded potential in DR for $d\to 3$. The Fourier transform of (\ref{expanded}) is
\bea
 V_{2 \,{\rm exp}}(r)=&&-\frac{g^4C_F N_c T}{16\pi m_D}\left\{\left[
\left(1-\frac{3 \pi ^2}{16}\right)\frac{1}{4 \pi r} -  \frac{ m_D}{ \pi} L(r)
- \left(\frac{5\pi ^2}{24}-\frac{4
   }{3}\right)\frac{r m_D^2}{8\pi}\right]\right. 
	\label{expanded-r}\\[2mm]
&& \left.     -i\pi T m_D \left[ 
\frac{7r}{3 \pi^2} -\left(1-\frac{3 \pi ^2}{16}\right) \frac{m_D r^2}{24\pi^2}\left(1-L(r)
\right) -\frac{1}{\pi^3 T}\left(1-\frac{N_f}{2N_c}\right) L(r)
\right]\right\}\nonumber\,
\eea
where
$L(r)=-2/(d-3)+\gamma+\ln\left[\pi (r\mu)^2\right]$. It is easy to see that the $1/(d-3)$ poles can be absorbed into the parameters of $V_{2\,{\rm soft}}$ in (\ref{soft}) using
\bea\label{renormalization}
q_0&=& -\frac{C_F N_c }{16\pi^2} L\left(\frac{1}{m_D}\right) +{\bar q}_0\nn\\
i_0&=& \frac{g C_F N_c }{16\pi^3}\left(1-\frac{N_f}{2N_c}\right) L\left(\frac{1}{m_D}\right) +{\bar i}_0 \nn \\
i_2 &=& -\frac{C_F N_c m_D\left(1-\frac{3 \pi ^2}{16}\right) }{384 \pi ^2 g T}  L\left(\frac{1}{m_D}\right) + {\bar i}_2
\eea
where ${\bar q}_0$, ${\bar i}_0$ and ${\bar i}_2$ are finite when $d\to 3$.
This gives
\bea
V_2(r)&=&V_{2\,{\rm soft}}+V_{2 \,{\rm exp}}(r)=g^4 {\bar q}_0 T 
+ i\left(g^3 {\bar i}_0 T+ g^5 {\bar i}_2 r^2 T^3\right) \nn\\[3mm]
 && -\frac{g^4C_F N_c T}{16\pi m_D}\left\{\left[
\left(1-\frac{3 \pi ^2}{16}\right)\frac{1}{4 \pi r} -  \frac{ m_D}{ \pi} \ln (r^2 m_D^2)
- \left(\frac{5\pi ^2}{24}-\frac{4
   }{3}\right)\frac{r m_D^2}{8\pi}\right]\right. 
	\label{expanded-r-again}\\[2mm]
&& \left.     -i\pi T m_D \left[ 
\frac{7r}{3 \pi^2} -\left(1-\frac{3 \pi ^2}{16}\right) \frac{m_D r^2}{24\pi^2}\left(1-\ln (r^2 m_D^2) 
\right) -\frac{1}{\pi^3 T}\left(1-\frac{N_f}{2N_c}\right) \ln (r^2 m_D^2)
\right]\right\}\nonumber\,.
\eea
The damped approximation, which gives the potential in eqs.~(\ref{mikko-real-2} - \ref{dampladder-t}), regulates the poles in \eqref{expanded-r}. Physically the calculation reshuffles part of the soft contribution into the semi-hard one. In fig.~\ref{plots-exp} we show the BLO potential and the corresponding expanded approximation to it, shifted so it matches the BLO potential at $r=0.01$~fm, for $g=1.8$ and $T=180$~MeV. The fact that a shift makes the two potentials agree at short distances is consistent with the form in eq.~(\ref{soft}).

Finally we return to the fact that in our calculations we have consistently dropped constant terms that correspond to contributions from the quark self-energies (see the discussion under eq.~(\ref{curly})). 
To calculate them consistently at the order we are working at would require HTL vertices in some diagrams and is thus prohibitively difficult. 
However, these contributions just 
add to the constant terms in (\ref{soft}), which means that when the coefficients $(q_0,i_0)$ are obtained by fitting to lattice data in sec. \ref{sec-lattice}, the quark self-energies are automatically included. 

\section{Perturbative computation of the spectrum}
\label{sec-pert}

The leading order coordinate space potential expanded in $r$ is 
\be
V_{\rm LO} (r) = - \frac{\alpha_s C_F}{ r}  -   \frac{\alpha_s C_F m^2_D}{2 }  { r} +  i  \frac {\alpha_s}{3}  C_f T (m_D r)^2 \left( \ln{\hat r} +  \gamma - \frac{4}{3}  \right) +\mathcal{O}\left(\alpha_s m_D^3 r^2\right)\,.
\label{H1lo}
\ee
In this section we will use the QCD  fine structure constant defined $\alpha_s = g^2/(4\pi)$. The largest term in (\ref{H1lo}) is the first term which is just the QCD Coulomb potential. 
The spectrum of this term is well known. 
The bound state energies and wavefunctions can be obtained from the solutions for the hydrogen atom in any quantum mechanics textbook, with appropriate rescalings of the coupling and the proton mass. 
We will define the Bohr radius $a= 2/( M_Q C_F \alpha_s)$ where $M_Q$ is the mass
of the heavy quark and use the dimensionless variable $\rho = 2 r/(a n)$ where $n$ is the principle quantum number. We always assume  that the mass of the heavy quarks that make up the quarkonium bound states is much greater than the temperature: $M_Q\gg T$.  The bound state energies are 
\be
E^{(0)}_n = -\frac 14  M_Q (C_F \alpha_s)^2 \frac{1}{n^2}  \,.
\label{E-lead}
\ee
The corresponding wavefunctions can likewise be obtained from any quantum textbook and depend on the principle quantum number $n$ and the quantum number $l$ which gives the eigenvalue of the angular momentum operator. These expressions are given in appendix \ref{Hatom}.

In this section we treat all corrections to the Coulomb potential perturbatively. This means we calculate corrections to the energy and the thermal width as
\bea
&& E' = \langle {\rm Re} [V-V_{\rm coulomb}]\rangle \\[1mm]
&& \Gamma = -2\langle {\rm Im} [V]\rangle
\eea
where the angle brackets indicate an expectation value calculated with the Coulomb wavefunction. 
Explicit expressions for the expectation values we need are given in appendix~\ref{Hatom}. 
The perturbative corrections to the energies and width from the non-Coulomb terms in (\ref{H1lo}) are
\bea
&& E^{\rm LO}_{n, \ell} = - \frac{ 2 \pi \alpha_s T^2 \hat m_D^2}{M_Q}  n  \langle \rho  \ \rangle_{n,\ell} \nn \\[3mm]
&& \Gamma^{\rm LO}_{n,\ell}  = - \frac{8 \pi \hat m_D^2 T^3}{3 C_F M^2_Q} n^2 \left[  \left(\ln \left(\frac{2 \sqrt{\pi} T \hat m_D n e^{\gamma_E}}{\sqrt{\alpha_s} C_F M_Q}\right) - \frac 43  \right) \langle \rho^2  \ \rangle_{n,\ell} +   \langle \rho^2 \ln \rho \ \rangle_{n,\ell} \right]\,.
\eea

To calculate the perturbative corrections from the BLO potential we start from the momentum space form. 
The reason is that BLO we will add fitted contributions of the form (\ref{soft}) and the form of these contributions is determined from the momentum space expression (as explained in sec.~\ref{sec-fit}). 
The expanded momentum space potential $V_{2\,{\rm exp}}(p)$ is given in eq.~(\ref{expanded}) and the  Fourier transform of this expression is eq.~(\ref{expanded-r}). 
We will add the contributions in eq.~(\ref{renormalization}) which removes the divergences and the dependence on the factorization scale, as in sec.~\ref{sec-fit}.
The results for the perturbative corrections from the BLO potential are
\bea
E^{\rm BLO}_{n, \ell} &=& 
\frac{\sqrt{\alpha_s} N_c}{4 \sqrt{ \pi} {\hat m_D}}  \left(1-  \frac{3 \pi^2}{16} \right) E_n^{(0)} 
+ 2 \alpha_s^2 N_c C_F T 
\left(
\ln\left(\frac{2 \sqrt{\pi} \hat m_D n T}{\sqrt{\alpha_s}C_F M_Q}\right) +
\left\langle \ln\left( \rho \right)\right\rangle_{n,\ell}
\right) \nonumber \\[2mm]
&+ & \frac{ \alpha_s^{3/2} N_c T^2 \sqrt{\pi} {\hat m_D}}{M_Q} \left( \frac{5 \pi^2}{96} - \frac13      \right) n   \langle \rho  \rangle_{n,\ell} + 16 \pi ^2  T \alpha _s^2\,\bar{q}_0
\label{eHatom} \\[4mm]
\Gamma^{\rm BLO}_{n,\ell}  &=& 
- \frac{14 \alpha_s N_c T^2}{ 3 M_Q} n \langle \rho \ \rangle_{n,\ell} \nn\\[2mm]
&+& \frac{\sqrt{\pi }  n^2  T^3 N_c \hat{m}_D \sqrt{\alpha _s}}{6 C_F M_Q^2}
\left(1-\frac{3 \pi ^2}{16}\right)
 \left[
\left\{1-2 \ln \left(\frac{2 \sqrt{\pi } n T \hat{m}_D}{C_F M_Q \sqrt{\alpha _s}}\right)\right\}\langle\rho^2\rangle_{n, \ell} - 2\langle \rho^2 \ln (\rho )\rangle_{n, \ell} 
 \right] \nn \\[2mm]
& +&   \frac{ 4 \alpha^2_s  C_FN_c T}{  \pi}  \left(1-\frac{N_f}{2N_c}\right)  
\left(\ln\left(\frac{2 \sqrt{\pi}  \hat m_D n T}{\sqrt{\alpha_s}C_F M_Q}\right) + \langle \ln {\rho} \rangle_{n, \ell}    \right) \nn \\[2mm]
& - & 16 \pi ^{3/2}  T \alpha _s^{3/2} \,\bar i_0 -\frac{64 \pi ^{5/2}  n^2 \rho ^2 T^3 \sqrt{\alpha _s}}{C_F^2 M_Q^2}\,\bar i_2\,.\label{gamma-pert-2}
 \eea

It should be added that heavy quarkonium bound states are not believed to be well described with a perturbative treatment beyond the ground state \cite{Leutwyler:1980tn}. This is because the size of the non-perturbative corrections grows rapidly with the principal quantum number. We observe a similar behavior with the thermal corrections in \eqref{eHatom} and \eqref{gamma-pert-2}. 
Hence, although higher-order perturbative calculations for excited states at zero temperature exist in the literature (see, for instance \cite{Brambilla:1999xj,Kiyo:2013aea,Anzai:2018eua}), we do not expect our perturbative results to provide a good description beyond the ground state.

\section{ Comparison with lattice data}
\label{sec-lattice}

Even though our calculation is valid in a very specific setting, we would like to test  how it
compares to lattice calculations which have a more general setup. We will first compare with the results of ref.~\cite{Bazavov:2023dci} for the static potential and next with those of ref.~\cite{Larsen:2019zqv} for the spectrum and decay width.
In all numerical calculations we use $N_c=N_f=3$. 

\subsection{Lattice potential}
\label{sec-rasmus}

\begin{figure}[H]
\begin{centering}
\includegraphics[scale=0.89]{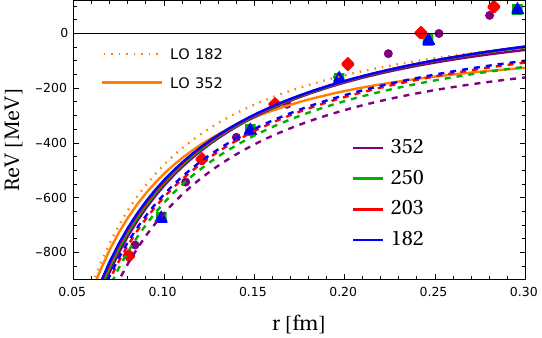}
\includegraphics[scale=0.88]{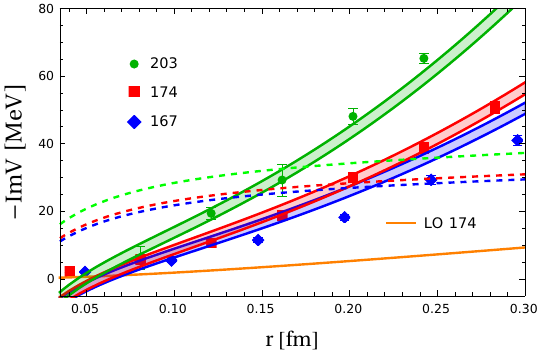}
\caption{Real and imaginary part of $V$. The dashed (solid) lines correspond to $V=V_{1{\rm LO}}+ V_2$ ($V= V_{1{\rm LO}}+ V_2+V_{2{\rm soft}}$) with parameters given in the text. The legends indicate the temperature in MeV. The real part includes a global shift of $S_{\rm ave}=209.5$~MeV for all curves. The LO results include the one-loop static quark self-energies. For the LO imaginary part
we show a single temperature since the other two would overlap with it.
 The solid bands on the plot of the imaginary potential indicate the uncertainty in the values of the fitted constants inherited from the error bars of the lattice data. \label{real-fit}}
\end{centering}
\end{figure}

First we compare our potential with the result of the lattice calculation in ref.~\cite{Bazavov:2023dci}. We fix $g$ from a fit to the $T=0$ lattice data for the static potential with $r\in [0,0.3]$ fm which delivers $g=1.8$.\footnote{The running of $g=g(T)$ is due to the power corrections to the real part of the potential \cite{Manuel:2016wqs,Carignano:2017ovz} which are higher order in our counting ($g^2$ suppressed). We also note that although using a value of $g$ greater than one appears to be inconsistent with the inequality $g\ll 1$ that we have used to structure our calculation, in fact our BLO potential does give a moderate correction to the LO potential, as can be seen in fig.~\ref{plots-ana}. This happens because the size of the corrections we calculate is determined by some smaller function of $g$, like $gN_c/(16\pi\hat m_D)\sim 0.049\, g$ (see eq.~(\ref{re20})).} The coefficients $(q_0,i_0,i_2)$ are determined by fitting to the lattice data at all available temperatures and values of $r$ between 0.04 and 0.3~fm.\footnote{In \cite{Carrington:2025cnv} we consider an alternate procedure where we select points using an upper bound on $rT$. The difference between the two methods is not significant.}  To adjust the origin of energies, we also add a temperature independent constant $S$ to the real part of the potential. 
To make a fair comparison we adjust the origin of energies for the LO potential, find $S_{\rm LO}$, and plot results using $S_{\rm avg}=(S+S_{\rm LO})/2$. 
The values we obtain are 
$(q_0,i_0,i_2)=(0.027,-0.019\pm 0.001, 0.194\pm 0.002)$ and $S_{\rm avg}=(S+S_{\rm LO})/2=209.5$ MeV.
The potentials are shown in fig.~\ref{real-fit}. If we ignore the soft contributions ($q_0=i_0=i_2=0$) our results are still closer to data than the LO result, although the improvement is marginal.  
When we include the soft contributions we observe that the real part of the potential varies very little with temperature, in agreement with the data. This is non-trivial since the soft contribution is independent of $r$. The slightly different shape with respect to the data is expected to be fixed by higher order temperature independent corrections, see ref.~\cite{Bazavov:2012ka}. For the imaginary part of the potential most of the BLO contribution comes from the soft region, which is responsible for the large correction with respect to the LO result. 

\subsection{Binding energies and widths}
\label{sec-peter}

To investigate the physical effects of the BLO corrections to the potential that we have calculated, we solve the Schr\"odinger equation. 
Since the potential is complex one should properly solve a set of coupled equations whose eigenvalues would give the binding energy and thermal width of each state. We make the approximation that we can separate the real and imaginary parts of the equation. We will use the real part of the potential to find the binding energy and the corresponding wavefunction, and then calculate the decay width of the state as the expectation value $\Gamma = -2\langle{\rm Im}[V]\rangle$. 

Assuming a spherically symmetric potential we can separate variables by writing the wavefunction in the form $\Psi(t,\vec r) = \varphi(t)Y_{l0}(\theta,\phi) R_{nl}(r)$ where the functions $Y_{lm}(\theta,\phi)$ are the standard spherical harmonics. We define $u_{nl}(r) = r R_{nl}(r)$ and the radial part of the reduced Schr\"odinger equation is
\bea
&& -\frac{\hbar^2}{2m} \left(\frac{d^2 u(r)}{dr^2} - \frac{l(l+1)}{r^2}u(r)\right) + V(r) u(r) = E u(r)\,.
\eea
We use $M_Q = 2m$ and define the scaled variables
\bea
&& \bar r = r/a \text{~~with~~} a=4\pi/(g^2 C_F M_Q) \nn \\
&& \hat V = M_Q a^2 V \text{~~and~~} \hat E = M_Q a^2 E \nn
\eea
so that the Schr\"odinger equation can be written in the form
\bea
-\frac{d^2 u(\bar r)}{d\bar r^2} + \frac{l(l+1)}{\bar r^2} u (\bar r) + \hat V(\bar r) u(\bar r) = \hat E u (\bar r) \,.\label{my-sch}
\eea
We solve this equation with the inital conditions $u(\bar r) = \bar r^{l+1}$ and
$u'(\bar r) = (l+1)\bar r^{l}$
using the method of ref.~\cite{Lucha:1998xc}. 
The basic strategy is as follows. 
The maximum of the potential is its value as $\bar r\to\infty$. The minimum is found numerically. The differential equation is then solved using a trial eigenvalue $E_{\rm try} = (E_{\rm min}+E_{\rm max})/2$. 
The functions in eq.~(\ref{meijerG}) are found by evaluating the integrals at small $\bar r$ and using an asymptotic series at large $\bar r$. 
If the solution we obtain is not normalizable we adjust the trial energy and try again. We know to adjust the trial energy either up or down based on whether the solution diverges to positive or negative infinity and if the number of nodes is even or odd. The true eigenvalue can be found quickly using a bilinear search method. 
In numerical calculations we consider only bottomonium and take $M_Q=4676$~MeV. 

\begin{figure}[H]
\begin{centering}
\includegraphics[scale=1.28]{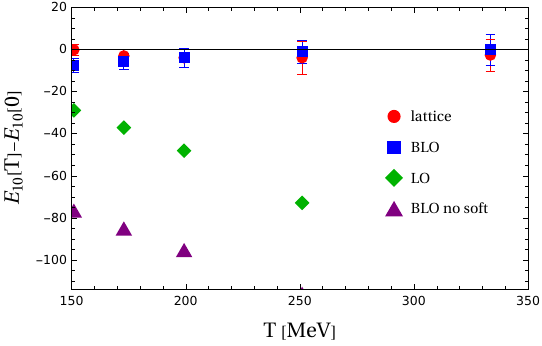}
\caption{The temperature dependence of the binding energy of the ground state. \label{EB-fit}}
\end{centering}
\end{figure}

We compare our results with those of the lattice calculation in ref.~\cite{Larsen:2019zqv}. 
First we discuss the binding energies. 
To make the comparison we include the lowest order contribution from the quark self-energies ($-g^2 C_F m_D/(4\pi)$) and subtract the vacuum (Coulomb) eigenvalue. To take into account the contributions to the potential from the soft momentum regime we add a term $q_0 g^4 T$, as explained in sec.~\ref{sec-fit}, and find the coefficient $q_0$ by fitting to the results of ref.~\cite{Larsen:2019zqv}. We find $q_0 = 0.044\pm 0.002$ where the error bar is obtained by fitting to the upper and lower lattice values. 
The results are shown in fig.~\ref{EB-fit}. If we do not include the soft contributions then the agreement of the BLO binding energies with the lattice results is slightly worse relative to the LO calculation. 

\begin{figure}[H]
\begin{centering}
\includegraphics[scale=1.18]{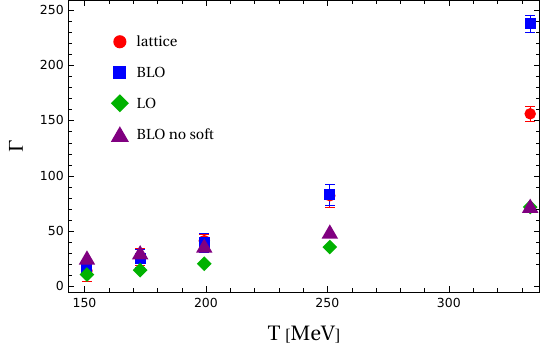}
\caption{The temperature dependence of the width of the ground state. \label{GAM-fit}}
\end{centering}
\end{figure}

We can also compare the thermal widths obtained from our calculation with the lattice results. 
To match the definition of ref. \cite{Larsen:2019zqv}
the widths for each temperature are calculated as the expectation value $-\langle {\rm Im}[V]\rangle$.
The soft contributions have the form $i_0 g^3 T+i_2 g^5 r^2 T^3$ (see section \ref{sec-fit}).  The constants $(i_0,i_2)$ are determined by fitting to the lattice data at all available temperatures. 
This gives $(i_0,i_2)=(-0.026 \pm 0.009 ,0.052 \pm 0.002)$, with errors obtained from fitting to the upper and lower lattice values. The results are shown in fig.~\ref{GAM-fit}. If the soft contributions are not included we still find a slight improvement in the description of the decay width relative to the LO result. 
Figures \ref{EB-fit} and \ref{GAM-fit} show that when the soft contributions are included  
the temperature dependence of both the binding energy and the decay width give a reasonable description of lattice data, and a considerable improvement with respect to the LO results. 

\subsection{Dissociation temperature}

\begin{figure}[H]
\begin{centering}
\includegraphics[scale=1.28]{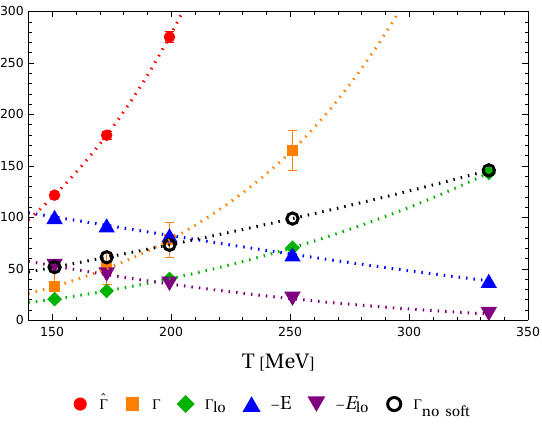}
\caption{
The binding energy ($-E$) and the thermal width ($\Gamma$) as a function of temperature. The dotted lines are interpolated curves that are added to guide the eye.  The notation 
$\Gamma_{\rm no\,\, soft}$, $\Gamma$ and $\hat\Gamma$ indicate the curves obtained from the BLO calculation without soft contributions, and with soft contribution using the fits to ref.~\cite{Larsen:2019zqv} 
and ref.~\cite{Bazavov:2023dci} 
respectively.   In all cases the contribution from the LO quark self-energies is included in the results for the widths. 
 \label{DIS-fit}}
\end{centering}
\end{figure}

We can use our results for the binding energies and widths to estimate the dissociation temperature. 
The dissociation temperature of a bound state can be defined as the temperature for which its thermal decay width equals the energy difference to the closest state, since this is the point at which we are not able to distinguish  a given state from its neighbour in the spectral function. For simplicity we will use instead the energy difference to the threshold, which is expected to provide a slightly higher dissociation temperature.
The decay width is defined to be 
the expectation value of $-2$Im$[V]$. The binding energies and widths as a function of the temperature are shown in fig.~\ref{DIS-fit}. We calculate the dissociation temperature by finding numerically the point where the curves for the binding energy and thermal width cross. The results are shown in table~\ref{Tdis}. 
We note that the dissociation temperature we obtain from the fit to ref.~\cite{Bazavov:2023dci} is very low, incompatible with earlier lattice studies that indicate it is much higher than the crossover temperature (see for instance \cite{Aarts:2014cda}).
This discrepancy might be due to a problem in the extraction of the decay width in the lattice calculation. 
\begin{table}[H]
\centering 
\begin{tabular}{|c | c | c|c|c|} 
\hline               
~~~~ & ~~~~~~~LO~~~~~~~~ & ~~~ BLO no soft ~~~ & ~~~ BLO fit \cite{Larsen:2019zqv}  ~~~& ~~~ BLO fit \cite{Bazavov:2023dci}  ~~~ \\ [0.5ex]
\hline
$T_{\rm diss}$ [MeV] &193.2 & 210 & $202\pm 10$ & $142.7\pm 1.1$\\
\hline
\end{tabular}
\caption{The dissociation temperature using different approximations (see text for details). \label{Tdis}}
\end{table}

\subsection{Analysis}

The numerical coefficients that we have obtained by fitting to two different sets of lattice data, as described in sections \ref{sec-rasmus} and \ref{sec-peter}, are shown in table~\ref{fitted-coefficients}. Ideally the values of the coefficients obtained from the two different sets of data would agree within uncertaintities. In addition, since all scales are explicit in our approach, the three coefficients should have approximately the same size. This is true for all coefficients in the two sets except for $i_2$ obtained from the data of ref.~\cite{Bazavov:2023dci}. The outcomes for $q_0$ differ by only a factor $\approx 0.61$, for $i_0$ by a factor $\approx 0.73$, but for $i_2$ they differ by a factor $\approx 3.7$. As mentioned above, the dissociation temperature obtained from the fit to ref.~\cite{Bazavov:2023dci} is unrealistically low. 
\begin{table}[H]
\centering 
\begin{tabular}{|c | c | c|c|c|} 
\hline                       
~~~~~~ & $~~~~~~~~~q_0~~~~~~~~~~$ & ~~~~~~~~~~ $i_0$ ~~~~~~~~~~ & ~~~~~~~~~~ $i_2$  ~~~~~~~~~~ & ~~~~~~~~~~ $S_{\rm ave}$    ~~~~~~~~~~ \\ [0.5ex]

\hline
\cite{Bazavov:2023dci} & 0.027 & -0.019 $\pm$ 0.001 & 0.194 $\pm$ 0.002 & 209.5 MeV \\
                                    
\hline

\cite{Larsen:2019zqv} & 0.044$\pm$ 0.002 & -0.026 $\pm$ 0.009 & 0.052$\pm$ 0.002 & - \\
\hline
\end{tabular}
\caption{The values of the  coefficients obtained from fits to different sets of lattice data. The first row corresponds to the fitting procedure used in sec.~\ref{sec-rasmus} and the second row is obtained in sec.~\ref{sec-peter}. \label{fitted-coefficients}}
\end{table}

\section{ Conclusions}
\label{sec-concl}

We have elaborated on a recent computational scheme for the real-time finite-temperature static potential presented in ref.~\cite{Carrington:2024ize}. It is designed to produce a good approximation to the actual potential in the region where bound states become wide resonances and when the coupling constant is small. We have discussed extensively all subtleties in the calculation, both in the main text and in the appendices. 
We have shown that soft contributions to the coordinate space potential at short distances have a universal form. We have used our potential to calculate
the heavy quarkonium thermal energy shift and thermal decay width.

We have pushed our computational scheme beyond its strict range of applicability and compared it to recent lattice data.
After fitting the universal soft contributions to data, we get a good description both of the potential of ref. \cite{Bazavov:2023dci}   and of the $\Upsilon (1s)$ thermal energy shift and thermal decay width of ref. \cite{Larsen:2019zqv}. We have used these results to estimate the dissociation temperature of the $\Upsilon (1s)$ state.

Finally, our results encourage us to propose our formulas for the real-time static potential as physically motivated inputs for the Bayesian methods which are mandatory in order to extract the potential non-perturbatively from lattice data. This is a notoriously difficult task. We recall that in the early attempts to identify the potential \cite{Kaczmarek:1999mm,Kaczmarek:2002mc,Petreczky:2004pz,Kaczmarek:2005ui,Maezawa:2007fc, Burnier:2009bk} the imaginary part was completely missed. We have provided two sets of formulas. The fully expanded ones (\ref{expanded-r}), which are strictly consistent with our power counting procedure, and the results corresponding to the damped approximation (\ref{mikko-real-2} - \ref{dampladder}), which partially 
resum some of the soft contributions.

\begin{acknowledgments}

We thank Peter Petreczky and Rasmus Larsen for providing the data of refs.~\cite{Larsen:2019zqv} and \cite{Bazavov:2023dci}, respectively. MEC acknowledges support by the Natural Sciences and Engineering Research Council of Canada under grant SAPIN-2023-00023 and thanks ICCUB and ICE for hospitality. CM  was supported by Ministerio de Ciencia, Investigaci\'on y Universidades (Spain) MCIN/AEI/10.13039/ 501100011033/ FEDER, UE, under the project  PID2022-139427NB-I00, by Generalitat de Catalunya by the project 2021-SGR-171 (Catalonia), and also partly supported by the Spanish program Unidad de Excelencia
 Maria de Maeztu CEX2020-001058-M, financed by MCIN/AEI/10.13039/501100011033.
JS acknowledges financial support from Grant No. 2021-SGR-249 from the Generalitat de Catalunya and from projects No. PID2022-136224NB-C21, PID2022-139427NB-I00, No. CEX2019-000918-M and No. CEX2024-001451-M from Ministerio de Ciencia, Innovaci\'on y Universidades (MICIU/AEI/10.13039/501100011033).

\end{acknowledgments}

\appendix

\section{Notation}
\label{appendix-notation}

The vector potential $A_\mu$ is an SU($N_c$) valued function that can be written as a linear combination of the generators $T_a$ that satisfy $[T^a,T^b]=i f_{abc}T_c$ and Tr($T_aT_b$) = $\delta_{ab}/2$. We define 
$C_F =  (N_c^2-1)/(2N_c)$. We use $g$ to denote the QCD coupling constant and $T$ is the temperature. The Debye mass is written $m_D$. We also use the notation $m_D = g T \hat m_D$ with $\hat m_D = \sqrt{(N_c+N_f/2)/3}$ where $N_f$ is the number of light flavours. We use the standard notation $r=|\vec r|$. In some places we use the dimensionless variables $\hat p = p/m_D$ and $\hat r = r m_D$. We use dimensional regularization (DR) and write $D=d+1$ with $d=3+2\epsilon$. 
We use capital letters for four-vectors and, for example, $P^2=p_0^2-p^2$.

In our notation the Dyson equation has the form 
\bea
G^{-1}_{\mu\nu} = G^{-1}_{\mu\nu} - \Pi_{\mu\nu}
\eea
and the bare inverse propagators in covariant and Coulomb gauge are
\bea
&& G^{-1}_{\mu\nu\text{~cov}} = P^2\left(g_{\mu\nu}-\frac{P_\mu P_\nu}{P^2}\right) + \frac{1}{\chi}P_\mu P_\nu \nn \\
&&G^{-1}_{\mu\nu\text{~cou}} = P^2\left(g_{\mu\nu}-\frac{P_\mu P_\nu}{P^2}\right) + \frac{1}{\chi}(P_\mu-p_0g_{\mu 0})(P_\nu-p_0g_{\nu 0}) \,.
\eea
To decompose the polarization tensor we use $n^\mu \equiv \left(g^{\mu\nu}-P^\mu P^\nu/P^2\right) g_{\nu 0}$ and define the projectors
\bea
\label{tensors-A-B}
&& P_1^{\mu\nu} \equiv g^{\mu\nu} - \frac{P^\mu P^\nu}{P^2} - \frac{n^\mu n^\nu}{n^2} \text{~~and~~}
 P_2^{\mu\nu} \equiv \frac{n^\mu n^\nu}{n^2}\,. \nn
\eea
We write the polarization tensor
\bea
&& \Pi^{\mu\nu}  = \Pi^T  \, P_1^{\mu\nu} + \Pi^L  \, P_2^{\mu\nu}  \nn \\
&& \Pi_L = -\Pi_{00}\frac{P^2}{p^2} \nn \\
&& \Pi_T = \frac{1}{D-2} (\Pi_\mu^{~\mu} + \frac{P^2}{p^2} \Pi_{00}  ) \,.
\eea
The propagator in covariant gauge is
\bea
&& G^{\mu\nu}_{\text{~cov}} = D_T P^{\mu\nu}_1 +D_L P^{\mu\nu}_2 +\chi \frac{P^\mu P^\nu}{P^4} \nn \\
&& D_T = \frac{1}{P^2-\Pi_T} \nn \\
&& D_L = \frac{1}{P^2-\Pi_L} =  \frac{p^2}{P^2(p^2+\Pi_{00})} 
\eea
and in strict Coulomb gauge ($\chi=0$)
\bea
&& G^{\mu\nu}_{\text{~cou}} = D_T P^{\mu\nu}_1 + D_{L\text{-cou}} n_0^\mu n_0^\nu \text{~~with~~} n_0^\mu = (1,0,0,0) \nn \\
&& D_T = \frac{1}{P^2-\Pi_T} \nn \\
&& D_{L\text{-cou}} = -\frac{P^2}{p^2(P^2-\Pi_L)} =  - \frac{1}{(p^2+\Pi_{00})} \,.
\eea

The 00 component of the HTL retarded propagator in Coloumb gauge is
\bea
[G_{00}^{\rm ret}]_{\rm htl} = -\frac{1}{\left(p^2+m_D^2 - \frac{p_0 m_D^2}{2 p} (\bar L-i\pi\theta[1-\hat p_0^2])\right)}\,
\label{ret-prop}
\eea
with
$
\bar L \equiv \ln[|\hat p_0+1|]-\ln[|\hat p_0-1|]
$.
The real part of the retarded and time-ordered propagators is 
\bea
{\rm Re}\big[(G_{00}^{\rm ret})_{\rm htl}\big]= 
- \frac{p^2+m_D^2 - \frac{1}{2} \bar L \hat{p}_0 m_D^2}
{\left(p^2+m_D^2 -\frac{1}{2} \bar L \hat{p}_0 m_D^2\right){}^2
+\frac{1}{4} \pi ^2 \hat{p}_0^2 m_D^4 \Theta \left(1-\hat{p}_0^2\right)}\,.
\eea
The cut and pole parts of the imaginary part of the retarded propagator are
\bea
&& i{\rm Im}\big[(G_{00}^{\rm ret})_{\rm htl}\big]_{\rm cut} = 
\frac{i \pi  \hat{p}_0 m_D^2 \Theta \left(1-\hat{p}_0^2\right)}{2 \left(\left(p^2+m_D^2 -\frac{1}{2} \hat{p}_0 \bar{L}
   m_D^2\right){}^2+\frac{1}{4} \pi ^2 \hat{p}_0^2 m_D^4\right)} \nn \\[3mm]
&& i{\rm Im}\big[(G_{00}^{\rm ret})_{\rm htl}\big]_{\rm pole} =
-\frac{i \pi  \omega _L \left(p^2-\omega^2 _L\right) }{p^2 \left(p^2+m_D^2-\omega_L^2\right)} 
   \left(\delta \left(p_0-\omega _L\right)-\delta \left(p_0+\omega _L\right)\right)\,
\eea   
where $\omega_L$ is the solution to the equation
\bea
0=-\frac{1}{2} m_D^2 \omega _L \ln \left(-\frac{\omega _L+p}{p-\omega _L}\right)+p m_D^2+p^3
\,.\nn
\eea

\section{Some integrals and limits}
\label{appendix-integrals}

An integral representation of the theta function is
\bea
\Theta(x_0-y_0) = \frac{1}{2\pi i}\int \frac{dq_0}{q_0- i\eta}\,e^{iq_0(x_0-y_0)}\,.
\label{theta-int}
\eea
We need the limit as $t\to\infty$ of the integral
\bea
\int_0^t dx_0 \int_0^{t} dy_0 \, e^{il_0(x_0-y_0)} = \frac{4}{l_0^2}\sin^2\left(\frac{tl_0}{2}\right)\,.
\label{D1}
\eea
To find the leading order result we use the formula
\bea
&& \delta(x) = \lim_{\epsilon\to 0}\frac{1}{\epsilon} f\left(\frac{x}{\epsilon}\right) \text{~~where ~~} \int^\infty_{-\infty} dy\,f(y)=1 \nn
\eea
which with $\epsilon=1/t$ gives
\bea
\lim_{t\to\infty} t\,f(tx) = \delta(x)\,. \nn
\eea
Two useful options are
\bea
&& f(y) = \frac{\sin^2(y)}{\pi y^2} ~~\to~~ \lim_{t\to\infty} \frac{4}{t p_0^2} \sin^2\left(\frac{p_0 t}{2}\right) = 2\pi\delta(p_0) \label{option1} \\[2mm]
&& f(y) = \frac{\sin(y)}{\pi y} ~~\to~~ \lim_{t\to\infty} \frac{2}{p_0} \sin\left(\frac{p_0 t}{2}\right) = 2\pi\delta(p_0) \,.\nn 
\eea
From equations  (\ref{D1}, \ref{option1}) we have
\bea
\lim_{t\to\infty} \int_0^t dx_0 \int_0^{t} dy_0 \, e^{il_0(x_0-y_0)}
 = t\,2\pi\delta(l_0) + X\, \label{X1}
\eea
where $X$ is a $t$ independent constant.  
To find $X$ we multiply both sides of (\ref{X1}) by $l_0$ to get
\bea
&& \lim_{t\to\infty}\frac{4}{l_0}\sin^2\left(\frac{tl_0}{2}\right) = \frac{2}{l_0}\lim_{t\to\infty}\left(1-\cos(tl_0)\right) = l_0 X\,. \nn 
\eea
We will need to calculate the limit as $t$ goes to infinity of the function
\bea
f(t) = \frac{1}{l_0^2}\big(1-\cos(l_0 t)\big) = \frac{1}{2l_0^2}\left(1-e^{il_0t}\right)+\frac{1}{2l_0^2}\left(1-e^{-il_0t}\right) \,.\nn 
\eea
We can regulate each term consistently with the prescription
\bea
\lim_{t\to\infty} f(t) &=& \frac{1}{2}\lim_{t\to\infty} \left(\frac{1}{(l_0+i\eta)^2}\left(1-e^{i(l_0+i\eta)t}\right)
+ \frac{1}{(l_0-i\eta)^2}\left(1-e^{-i(l_0-i\eta)t}\right) \right) \nn \\[4mm]
&=& \frac{1}{2} \left(\frac{1}{(l_0+i\eta)^2}
+ \frac{1}{(l_0-i\eta)^2} \right)\,. \label{X-int}
\eea
The final result is therefore
\bea
\lim_{t\to\infty} \int_0^t dx_0 \int_0^{t} dy_0 \, e^{il_0(x_0-y_0)}
 = t\,2\pi\delta(l_0) + \left(\frac{1}{(l_0+i\eta)^2} +  \frac{1}{(l_0-i\eta)^2}\right)\,. \label{X4}
\eea

\section{ Corrections to the HTL gluon self-energy in Feynman gauge}
\label{appendix-feyn}

In this section we discuss the gauge independence of the part of the BLO potential from corrections to the HTL gluon self-energy. 
In sections \ref{sec-tony} and \ref{sec-pow} we calculated two different self-energy corrections in Coulomb gauge. We have called the first of these the HTL correction and the second is the POW. 
The POW is not gauge invariant, but the leading order contribution to the imaginary part in the limit $p_0\to 0$ is the same in both Coulomb and Feynman gauges, which is enough to ensure that the corresponding contribution to the potential is the same in both gauges. This is shown in sec.~\ref{boson-pow}. The HTL correction to the self-energy is not the same in Coulomb and Feynman gauges, even at leading order in the limit $p_0\to 0$, and thus the corresponding contribution to the potential is also not the same. However there is another contribution to the potential in Feynman gauge, at the same order, which is missing in Coulomb gauge. In sec.~\ref{sevenOthree} we calculate this extra contribution and show that the full Feynman gauge result agrees with the HTL corrected Coloumb expression. 

\subsection{Boson POW in Feynman gauge}
\label{boson-pow}

In Feynman gauge  the contribution of the gauge degrees of freedom to the POW is \cite{Ekstedt:2023anj,Gorda:2023zwy}
\begin{equation}
\Pi^{ \rm POW }_g (p_0, p)=   \frac{g^2 N_c}{48 \pi^2} p^2 \left[ 5\left( \frac 1\epsilon - \ln{\frac{\mu^2 e^\gamma}{T^2 4 \pi}} \right) - 2 \frac{p_0^2}{p^2} +  1 
+ \frac {p_0 ( p_0^2-6 p^2)}{p^3} \ln\left({ \frac{p_0+p+i\eta}{p_0-p+i\eta} }\right) \right] \,.
\end{equation}
It is easy to see that in the limit $p_0\to 0$ the imaginary part agrees with the Coulomb gauge result in eq.~(\ref{IM-POW}). As discussed in sec.~\ref{sec-count} the real part does not give a contribution to the potential at the order we are working at. 

\subsection{HTL corrected self-energy in Feynman gauge}
\label{sevenOthree}

In sec.~\ref{sec-tony} we calculated corrections to the HTL self-energy from the momentum region $m_D \ll p \ll T$ in Coulomb gauge. We can follow the same procedure to get the result in Feynman gauge and find
\bea
[\Pi_{00} (p_0\to 0, p)]_{\rm semi-hard} = -\frac{g^2 N_c T }{4} \left( p  + \frac{ 3 i  }{\pi} p_0 \right) + {\cal O}(p_0^2)\,. \label{tony-result-fyn}
\eea
Comparing with eq.~(\ref{tony-result}) we see that the real part is the same but the imaginary part is not. 
When we substitute (\ref{tony-result-fyn}) into the time-ordered propagator and expand in $p/T$ we find the contribution to the potential is 
\bea
&& \tilde V^{\rm Feyn}_{1{\rm HTL}}(p) = g^2 C_F \, G_{\rm nlo}(0,p) \nn \\[2mm]
&& G_{\rm nlo}(0,p)  = -\frac{g^2 N_c p\, T}{4(p^2+m_D^2)^2}   -\frac{i g^2 N_c \,3\,T^2}{2\pi(p^2+m_D^2)^2} \,\label{prop-HTL-fyn}
\eea
which does not agree with the Coulomb gauge result in eq.~(\ref{prop-HTL}). 
To resolve the discrepancy  we note that in Feynman gauge there is a contribution to the imaginary part of the potential from the ladder and crossed-ladder graphs in equation (\ref{ladd-cross-all}) in the regime where $k$ is semi-hard. This can be calculated using bare propagators. The integrals we need to calculate have the form
\bea
 \tilde V_2^{bc}(r) &=& -i\frac{g^4}{4}(N_c^2-1) \int \frac{d^dp}{(2\pi)^d} e^{i \vec p \cdot\vec r} X \nn\\[2mm]
X &=&   \int \frac{d^Dk}{(2\pi)^D}
\frac{1}{(k_0 + i\eta)^2}  [G_{00}(k_0,\vec k)G_{00}(k_0,\vec k+\vec p)]]_{g=0}\,. \label{sq-bracket}
\eea
The leading contribution to the integral $X$ comes from taking the imaginary parts of both propagators which gives a double Bose enhancement. 
We use equation (\ref{timeordered}) with $G^{\rm ret/adv}(p_0,p) = 1/(p^2_0-p^2\pm i\eta)$ and do the $k_0$ integral to obtain 
\bea
X &=& -\pi T^2 \int \frac{d^d\Omega}{(2\pi)^d} \int dk k^{d-6} \frac{1}{q}\big(\delta(k-q) + \delta(k+q)\big) \nn \\[2mm]
 &=& -2\pi T^2 \int \frac{d^d\Omega}{(2\pi)^d} \int dk k^{d-6} \delta(k^2-q^2) \nn 
\eea
where we have defined $q=\sqrt{p^2+k^2+2pk y}$ with $y=\hat p\cdot\hat k$ and 
\bea
\int\frac{d\Omega_d}{(2\pi)^d} = \frac{4}{(\sqrt{4\pi})^{d+1}\Gamma\left(\frac{d-1}{2}\right)} \int_{-1}^1 dy\, (1-y^2)^{\frac{d-3}{2}}\,.\nn
\eea
We rewrite the delta function as
\bea
\delta(k^2-q^2) = \frac{1}{2pk}\,\delta\left(y-\frac{p}{2k}\right)\,\nn
\eea
and use this delta to do the $y$ integral. 
The result is
\bea
X = -\frac{\pi T^2}{p}\,  \frac{4}{(\sqrt{4\pi})^{d+1}\Gamma\left(\frac{d-1}{2}\right)} \int_{p/2}^\infty dk\,k^{d-7}\left(1-\frac{p^2}{4k^2}\right)^{\frac{d-3}{2}}\,. \label{tempa}
\eea
The $k$ integral can be rewritten using $\hat k = 2k/p$ to obtain
\bea
I_k \equiv \int_{p/2}^\infty dk\,k^{d-7}\left(1-\frac{p^2}{4k^2}\right)^{\frac{d-3}{2}} = \left(\frac{p}{2}\right)^{d-6} \int^\infty_1 d\hat k \hat k^{-4} (\hat k^2-1)^{\frac{d-3}{2}} \,.\nn 
\eea
This integral is defined for $1<d<6$  and gives 
\bea
I_k = \frac{2^{7-d} p^{d-6} \Gamma \left(3-\frac{d}{2}\right) \Gamma
   \left(\frac{d-1}{2}\right)}{3\sqrt{\pi }}\,. \label{kint-res}
\eea
We substitute (\ref{kint-res}) into (\ref{tempa}) and write $d=3+2\epsilon$ and expand to obtain
\bea
X = -\frac{2T^2}{3p^4 \pi} \,.
\eea
The final result is 
\bea
i{\rm Im}[V^{bc}_{2\text{bare}}] &=& -\frac{i g^4  N_c\,C_F}{2} \int \frac{d^3p}{(2\pi)^3} e^{i \vec p \cdot\vec r} X \label{bare-ladd} \\[2mm]
&=& \frac{i g^4 N_c\,C_F  T^2}{3\pi}\, \int \frac{d^3p}{(2\pi)^3} e^{i \vec p \cdot\vec r} \frac{1}{p^4}
= - \frac{i g^4  N_c\,C_F  T^2 }{2\pi} \left(-\frac{2}{3}\right) \, \int \frac{d^3p}{(2\pi)^3} e^{i \vec p \cdot\vec r} \frac{1}{p^4} \,. \nn
\eea
It is straightforward to check that there is no corresponding contribution from the vertex and wavefunction renormalization graphs (see eq.~(\ref{vert-wave-all})), and that there is no real contribution to $V_{2\rm{bare}}^{bc}$.
The previous results (\ref{prop-HTL}, \ref{prop-HTL-fyn}) can be written in the form
\bea
i{\rm Im}[\tilde V_{1\text{HTL}}] = -  \frac{i g^4 N_c\, C_F   {\cal N}T^2}{2\pi}  \int \frac{d^3p}{(2\pi)^3} e^{i \vec p\cdot\vec r} \frac{1}{(p^2+m_D^2)^2} 
\eea
where ${\cal N}=3$ in Feynman gauge and $7/3$ in Coulomb gauge. 
If we add the contribution (\ref{bare-ladd}) to the Feynman gauge result we find that Feynman and Coulomb gauges agree to the order to which we are working.

\section{Pinch singularities}
\label{appendix-pinch}
In this appendix we show how to calculate the different contributions in eq.~(\ref{curly}) and how to prove that the result is free of pinch singularities. 

The second and third diagrams in fig.~\ref{fig-diag} are obtained from eq.~(\ref{22}) by choosing different contractions of the fields. The contraction that gives the graph in fig.~\ref{fig-diag}b is the one that connects the two potentials with coordinates $(x_0,\vec r)$ and $(w_0,0)$, and the two with coordinates $(z_0,0)$ and $(y_0,\vec r)$. The graph in fig.~\ref{fig-diag}c is obtained from the contraction of the potentials with coordinates $(x_0,\vec r)$ and $(z_0,0)$, and the two with coordinates $(w_0,0)$ and $(y_0,\vec r)$. The first set of contractions gives
\bea
\frac{i}{t} C_{2b}(t) && = -i\frac{g^4 C_F^2}{t} \int_0^t dx_0 \int_0^t dy_0 \int_0^t dz_0 \int_0^t dw_0
\int\frac{d^4l}{(2\pi)^4}  \int\frac{d^4k}{(2\pi)^4} \nn \\[4mm]
&& G(l_0,l) G(k_0,k) 
 \Theta(x_0-y_0)\Theta(w_0-z_0)
e^{-il_0(x_0-w_0)} e^{i\vec l \cdot \vec r}
e^{-ik_0(z_0-y_0)}e^{-i\vec k \cdot \vec r}\,. \label{ladd-alla}
\eea
To construct a pinch free contribution we add the second term in the first curly bracket in (\ref{curly}) using the integral form in (\ref{oneC1a}). The integrand has the same form as eq.~(\ref{ladd-alla}) except for the theta functions. The two pieces can be combined to give  
\bea
V_{2b} &=& \lim_{t\to\infty} 
\left\{\frac{i C_{2b}(t,\vec r)}{t} - \frac{i C^2_{1a}(t,\vec r)}{2t}\right\} \nn \\[1mm]
 & =& -i g^4 C_F^2 \lim_{t\to\infty}  \frac{1}{t}
   \int_0^t dx_0 \int_0^t dy_0 \int_0^t dz_0 \int_0^t dw_0
\int\frac{d^4l}{(2\pi)^4}  \int\frac{d^4k}{(2\pi)^4}  \label{ladd-all2} \\[3mm]
&& G(l_0,l) G(k_0,k) 
 \left[\Theta(x_0-y_0)\Theta(w_0-z_0)-\frac{1}{2}\right]
e^{-il_0(x_0-w_0)} e^{i\vec l \cdot \vec r}
e^{-ik_0(z_0-y_0)}e^{-i\vec k \cdot \vec r}\,.\nn
\eea
The combination of theta functions in the square bracket can be rewritten by making some changes of variable and using the fact that the time-ordered propagator is even when the frequency changes sign. The result of these manipulations is
\bea
V_{2b} & =& ig^4 C_F^2  \lim_{t\to\infty}\frac{1}{t}
   \int_0^t dx_0 \int_0^t dy_0 \int_0^t dz_0 \int_0^t dw_0
\int\frac{d^4l}{(2\pi)^4}  \int\frac{d^4k}{(2\pi)^4}  \label{ladd-all2b} \\[4mm]
&& G(l_0,l) G(k_0,k) 
 \Theta(x_0-y_0)\Theta(z_0-w_0)
e^{-il_0(x_0-w_0)} e^{i\vec l \cdot \vec r}
e^{-ik_0(z_0-y_0)}e^{-i\vec k \cdot \vec r}\,.\nn
\eea
Finally we use the integral representation of the theta function (\ref{theta-int}) and perform the integrals with the help of eq.~(\ref{X4}). The result is
\bea
V_{2b}= - i g^4 C_F^2 \int \frac{d^3 l }{(2\pi)^3}\int \frac{d^4k}{(2\pi)^4} \frac{1}{(k_0+i\eta)^2}
G(k_0,\vec k+\vec p) G(k_0,\vec k)  e^{i \vec p \cdot\vec r} \,\label{ladd-all}
\eea
and is free of pinch singularities. 

The second set of contractions does not give a pinch singularity and one obtains immediately the contribution from diagram \ref{fig-diag}c which is
\bea
V_{2c}  = -\frac{ig^4 C_F}{2N_c} 
\int\frac{d^3l}{(2\pi)^3}  \int\frac{d^4k}{(2\pi)^4} 
\frac{1}{(k_0 + i\eta)^2}
G(k_0,\vec k+\vec p) G(k_0,\vec k) 
e^{i \vec p \cdot \vec r}\,.\nn \\ \label{cross-all}
\eea
Combining eqs.~(\ref{ladd-all}, \ref{cross-all}) gives the result for $V_2^{bc}$ in eq.~(\ref{ladd-cross-all}). 

The fourth and fifth diagrams in fig.~\ref{fig-diag} are obtained from eq.~(\ref{31}) by choosing different contractions of the fields. The contraction that gives the graph in fig.~\ref{fig-diag}d is the one that connects the two potentials with coordinates $(x_0,\vec r)$ and $(y_0,\vec r)$, and the two with coordinates $(u_0,\vec r)$ and $(z_0,0)$. It is straightforward to show that doing the time integrals, and then using the delta functions that are produced, gives the pinch free result 
\bea
V_{2d}&=&
\frac{i g^4 C_F}{N_c}\int \frac{d^3p}{(2\pi)^3} e^{i \vec p \cdot\vec r} G(0,\vec p) \int \frac{d^4k}{(2\pi)^4}
\left(\frac{1}{k_0 + i\epsilon}\right)^2  G(k_0,\vec k)\,
\label{vert-int-2-do}
\eea
where we have included a factor of 2 that comes from including the contribution with 1 potential on the upper line and 3 on the lower line, instead of the reverse as in eq.~(\ref{31}).

The graph in fig.~\ref{fig-diag}e is obtained from eq.~(\ref{31}) with the contraction of the two potentials with coordinates $(x_0,\vec r)$ and $(u_0,\vec r)$, and the two with coordinates $(z_0,0)$ and $(y_0,\vec r)$. In this case there is an additional factor of 4, two from the contraction that puts the self-energy correction on the left of the vertical rung, and another factor of two from the contributions from taking 1 potential on the upper line and 3 on the lower line. 
Including this factor of 4 we have
\bea
 \frac{i}{t}C_{2e}(t) && = 
 \frac{4ig^4C_F^2}{t} \int_0^t dx_0 \int_0^t du_0 \int_0^t dy_0 \int_0^t dz_0
\Theta(x_0-u_0)\Theta(u_0-y_0) \nn \\[2mm]
&& \int\frac{d^4k}{(2\pi)^4}\int\frac{d^4l}{(2\pi)^4}G(k_0,k) G(l_0,l) e^{i\vec l\cdot \vec r}
e^{-i k_0(x_0-u_0)}e^{-i l_0(y_0-z_0)}
\,.\nn
\eea
To obtain a pinch free expression we need to include both terms in the last curly bracket in eq.~(\ref{curly}). Using  the integral forms in (\ref{oneC1a}, \ref{oneC1b}) and combining terms we find
\bea
V_{2e} &=& \lim_{t\to\infty} 
\left\{\frac{i C_{2e}(t,\vec r)}{t} - \frac{i C_{1a}(t,\vec r)C_{1b}(t)}{t}\right\} \nn \\[2mm]
 &=& 2i g^4 C_F^2 \lim_{t\to\infty}  \frac{1}{t} \int_0^t dx_0 \int_0^t du_0 \int_0^t dy_0 \int_0^t dz_0
 \int\frac{d^4k}{(2\pi)^4}\int\frac{d^4l}{(2\pi)^4}G(k_0,k) G(l_0,l) e^{i\vec l\cdot \vec r} \nn \\[4mm]
&& e^{-i k_0(x_0-u_0)}e^{-i l_0(y_0-z_0)}
\big(\Theta(u_0-y_0) - \Theta(y_0-u_0)\big)\Theta(x_0-u_0)\,.
\eea
To do these integrals we rewrite the second product of theta functions as
\bea
\Theta(y_0-u_0)\Theta(x_0-u_0) = \Theta(y_0-x_0)\Theta(x_0-u_0) + \Theta(x_0-y_0)\Theta(y_0-u_0)\,.
\eea
Once again we use the integral representation of the theta function (\ref{theta-int}) and perform the integrals with the help of eq.~(\ref{X4}). The result is 
\bea
V_{2e} = 2i g^4 C_F^2 \int\frac{d^3p}{(2\pi)^3} G(0,p) e^{i\vec p\cdot \vec r} \int\frac{d^4k}{(2\pi)^4}\left(\frac{1}{k_0+i\eta}\right)^2 G(k_0,k)  \,.
\label{curly-wave-done}
\eea
Adding the expressions in eqs.~(\ref{vert-int-2-do}, \ref{curly-wave-done}) we obtain the result in (\ref{vert-wave-all}). 

\section{Orders of neglected contributions}
\label{sec-orders}
In this appendix we explain how to get the results in eq.~(\ref{dropped-this}). 

The order of the HTL vertex corrections can be found by multiplying the largest self-energy corrections by the factor $m_D^2/p^2\sim g^{2-2a}$. From (\ref{retony}) the largest self-energy contribution to the real part of the potential is $\sim g^{4-3a}$, and comparing (\ref{imtony}, \ref{impow})  the largest self-energy contribution to the imaginary part is $\sim g^{4-4a}$. Multiplying by $g^{2-2a}$ gives that the HTL vertex corrections for the real (imaginary) parts are of order $g^{6-5a}$ ($g^{6-6a}$), as stated in eq.~(\ref{dropped-this}). 

\begin{figure}[H]
\begin{center}
\includegraphics[width=0.35\textwidth]{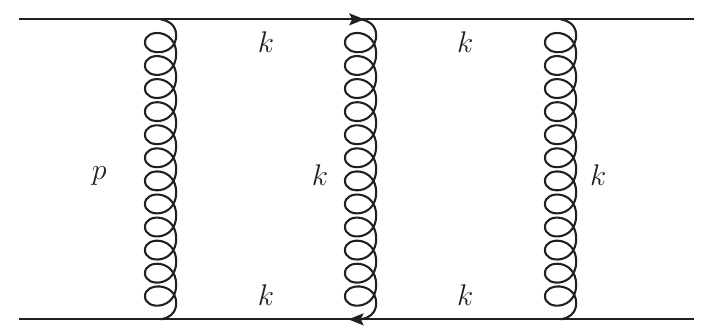}
\caption{An example of a two loop graph.  The semi-hard momentum flows through one of the gluon lines (which is labelled $p$) and all other lines are soft (labelled $k$). 
\label{triple-ladd}}
\end{center}
\end{figure}

It is more difficult to determine the order of the two loop graphs. 
As an example we consider the two loop ladder graph in fig.~\ref{triple-ladd}. 
There are an even number of quark propagators and vertices so these factors cannot introduce an overall factor of $i$. There are three gluon lines, which give a factor $\sim i^3$, and an additional overall $i$ from the factor $i/t$ in the definition of the potential (see eq.~(\ref{bigC})). Combining we find that there is no overall $i$. The product of the four quark propagators is also real for the graph in fig.~\ref{triple-ladd}.

The four quark propagators have momenta $\sim k$ so their contribution, together with the phase space factor from one of the loops gives $d^4k/k^4\sim 1$. 
The semi-hard momentum passes through one of the gluon lines (labelled $p$ in the figure) and the two remaining gluon propagators are soft (labelled $k$). 

We consider first the case when the semi-hard gluon propagator is real and $\sim p^{-2}\sim g^{-2a}\,T^{-2}$ (see eq.~(\ref{static-htl})). If they were real, the two soft gluon propagators together with the phase space factor for the second loop would give a factor $\sim d^4k/k^4 \sim 1$. 
The dominant contribution to the real part of the diagram will come from both of the soft gluon propagators imaginary, which gives an enhancement $\sim(T/k)^2\sim g^{-2}$. Combining with the overall factor from the vertices we have altogether $g^6 \times g^{-2a} \times g^{-2} = g^{4-2a}$. For the imaginary part, one of the soft gluon propagators must be real, which gives a relative factor $g$ to produce $g^{5-2a}$. 

Now we consider the case where the semi-hard gluon propagator is imaginary. From eq.~(\ref{static-htl}) this gives a factor $g^{2-5a}$ (instead of $g^{-2a}$). To make an overall real contribution one of the soft gluon propagators must be real instead of imaginary, which gives $g^{-1}$ (instead of $g^{-2}$). The relative factor is therefore $g^{2-5a}/g^{-2a} \times g^{-1}/g^{-2} = g^{3-3a}$ which gives for the real part $g^{4-2a}g^{3-3a}=g^{7-5a}$. For the imaginary part the real soft propagator is replaced by the imaginary soft propagator, so the relative factor is $g^{2-5a}/g^{-2a} \times g^{-2}/g^{-1} = g^{1-3a}$ which gives $g^{5-2a}g^{1-3a}=g^{6-5a}$. 

Combining we have that for a real (imaginary) semi-hard propagator the orders are
$g^{4-2a}$ ($g^{7-5a}$) for the real part and 
$g^{5-2a}$ ($g^{6-5a}$) for the imaginary part. 
Taking the largest result for $1/2<a<2/3$ gives that the real part is $\sim g^{4-2a}$ and the imaginary part is $\sim g^{5-2a}$. 

For two loop diagrams with different topologies the numbers of vertices and propagators (both gluon and fermion) is the same but the orientation of the lines is different. This means that there might be changes in the separation into real and imaginary parts described above. For this reason we make a conservative choice and assume that both the real and imaginary parts of the two loop diagrams are of order $g^{4-2a}$ (as stated in eq.~(\ref{dropped-this})).

\section{Expectation values}
\label{Hatom}

In this section we give the results for the expectation values needed 
in sec. \ref{sec-pert}. We define
\bea
\langle f(\rho)\rangle_{n,l} = \left(\frac{n}{2}\right)^3 \int_0^\infty d\rho \, \rho^2 \,f(\rho) \,\Psi^2_{nl}(\rho)
\label{integral}
\eea
where $\Psi^2_{nl}(\rho)$ is the hydrogen-like wavefunction of the QCD Coulomb potential written in terms of the dimensionless variable $\rho = 2r/(an)$ where $a$ is the Bohr radius (see section \ref{sec-pert}). This wavefunction is
\bea
\Psi_{nl}(\rho) = \frac{2  \sqrt{(-l+n-1)!} \,  \Gamma (l+n+1) }{n^2 \, ((l+n)!)^{3/2}}\, 
e^{-\rho /2} \, \rho ^l \, L_{-l+n-1}^{2l+1}(\rho )
\eea
where $L$ indicates an associated Laguerre polynomial defined using the normalization
\bea
\int_0^\infty d\rho\, e^{-\rho} \rho^j\,L_n^j(\rho)L_m^j(\rho) = \frac{(n+j)!}{n!}\delta_{nm} \,. \nonumber
\eea 
The factor in front of the integral in (\ref{integral}) comes from the Jacobian that converts the integral  over the radial variable to an integral over $\rho$.
It is straightforward to calculate the integral (\ref{integral}) when $f(\rho)$ is a polynomial.   Some additional tricks are needed when the function involves a logarithm \cite{Titard:1993nn}. The results we need are
\bea
\langle \rho^2  \ \rangle_{n,\ell} &=& 2 ( 5 n^2 - 3 \ell (\ell +1)+1) \  \\[2mm]
\langle \rho \ \rangle_{n,\ell}  &=& \frac{( 3 n^2 -  \ell (\ell +1))}{n} \\
\langle \rho^{-1} \ \rangle_{n,\ell}  &=& \frac{1}{2n} \label{rho-inverse}\\[2mm]
\langle \rho^2 \ln \rho \ \rangle_{n,\ell} &=&  \frac {1}{2n} \frac{\Gamma[{n-\ell}]}{\Gamma[{n + \ell +1}]} \times
\\
\nonumber
\sum_{j=0}^{n-\ell-1}&& \frac{12  \Gamma[ 5 + 2 \ell + j] \left(11 - 3 \gamma_E + 3 {\rm HNumber}[4 + 2 \ell + j] - 
   6 {\rm HNumber}[4 + \ell- n + j]\right))}{
\Gamma[-\ell + n - j]^2 \Gamma[1 + j] \Gamma[5 + \ell - n + j]^2} \\[2mm]
\langle \ln {\rho} \rangle_{n, \ell}  &=& \frac {1}{2n} \frac{\Gamma[{n-\ell}]}{\Gamma[{n + \ell +1}]} \sum_{j=0}^{n-\ell-1} \left( \frac{ 2 (1 - \gamma_E) \Gamma[3 + 2 \ell + j] }{\Gamma[-\ell + n - j]^2 \Gamma[1 + j] \Gamma[3 + \ell - n + j]^2} \right. \\
&+& \left. \frac{ \Gamma[3 + 2 \ell + j] \psi^{(0)}[ 3 + 2 \ell + j] -2 \Gamma[3 + 2 \ell + j] \psi^{(0)}[ 3 + \ell - n + j]
}{\Gamma[-\ell + n - j]^2 \Gamma[1 + j] \Gamma[3 + \ell - n + j]^2} \right)
\nonumber
\eea
where ${\rm HNumber}$ is the HarmonicNumber and $\psi^{(0)}(z)$ is the logarithmic derivative of the Gamma function.

{}


\begin{thebibliography}{}

\bibitem{Matsui:1986dk}
T.~Matsui and H.~Satz,
Phys. Lett. B \textbf{178}, 416 (1986).

\bibitem{Laine:2006ns}
M.~Laine, O.~Philipsen, P.~Romatschke and M.~Tassler,
JHEP \textbf{03}, 054 (2007).

\bibitem{Escobedo:2008sy}
M.~A.~Escobedo and J.~Soto,
Phys. Rev. A \textbf{78}, 032520 (2008).

\bibitem{ALICE:2020wwx}
S.~Acharya \textit{et al.} [ALICE],
Phys. Lett. B \textbf{822}, 136579 (2021).

\bibitem{CMS:2018zza}
A.~M.~Sirunyan \textit{et al.} [CMS],
Phys. Lett. B \textbf{790}, 270 (2019).

\bibitem{CMS:2017uuv}
A.~M.~Sirunyan \textit{et al.} [CMS],
Eur. Phys. J. C \textbf{78}, 509 (2018).

\bibitem{Carrington:2024ize}
M.~E.~Carrington, C.~Manuel and J.~Soto,
Phys. Rev. Lett. \textbf{134}, 011905 (2025). 

\bibitem{Beneke:1997zp}
M.~Beneke and V.~A.~Smirnov,
Nucl. Phys. B \textbf{522}, 321 (1998)

\bibitem{Smirnov:2012gma}
V.~A.~Smirnov,
Springer Tracts Mod. Phys. \textbf{250}, 1 (2012)


\bibitem{Jarrell:1996rrw}
M.~Jarrell and J.~E.~Gubernatis,
Phys. Rept. \textbf{269}, 133 (1996).

\bibitem{Asakawa:2000tr}
M.~Asakawa, T.~Hatsuda and Y.~Nakahara,
Prog. Part. Nucl. Phys. \textbf{46}, 459 (2001).

\bibitem{Rothkopf:2011ef}
A.~Rothkopf,
J. Comput. Phys. \textbf{238}, 106 (2013).

\bibitem{Burnier:2013fca}
Y.~Burnier and A.~Rothkopf,
Phys. Rev. D \textbf{87}, 114019 (2013).

\bibitem{Burnier:2013nla}
Y.~Burnier and A.~Rothkopf,
Phys. Rev. Lett. \textbf{111}, 182003 (2013).

\bibitem{Rothkopf:2011db}
A.~Rothkopf, T.~Hatsuda and S.~Sasaki,
Phys. Rev. Lett. \textbf{108}, 162001 (2012).

\bibitem{Burnier:2014ssa}
Y.~Burnier, O.~Kaczmarek and A.~Rothkopf,
Phys. Rev. Lett. \textbf{114}, 082001 (2015).

\bibitem{Burnier:2015tda}
Y.~Burnier, O.~Kaczmarek and A.~Rothkopf,
JHEP \textbf{12}, 101 (2015).

\bibitem{Burnier:2016mxc}
Y.~Burnier and A.~Rothkopf,
Phys. Rev. D \textbf{95}, 054511 (2017).

\bibitem{Lehmann:2020fjt}
A.~Lehmann and A.~Rothkopf,
JHEP \textbf{07}, 067 (2021).

\bibitem{Boguslavski:2021zga}
K.~Boguslavski, B.~S.~Kasmaei and M.~Strickland,
JHEP \textbf{10}, 083 (2021).

\bibitem{Bala:2021fkm}
D.~Bala \textit{et al.} [HotQCD],
Phys. Rev. D \textbf{105}, 054513 (2022).

\bibitem{Dong:2022mbo}
L.~Dong, Y.~Guo, A.~Islam, A.~Rothkopf and M.~Strickland,
JHEP \textbf{09}, 200 (2022).

\bibitem{Bazavov:2023dci}
A.~Bazavov, D.~Hoying, O.~Kaczmarek, R.~N.~Larsen, S.~Mukherjee, P.~Petreczky, A.~Rothkopf and J.~H.~Weber,
Phys. Rev. D {\bf 109}, 074504 (2024).

\bibitem{Zhou-CTP}
G.~Zhou, Z.~Su, B.~Hao and L.~Yu, Phys. Rep. 118, 1 (1985).

\bibitem{Ghiglieri:2020dpq}
J.~Ghiglieri, A.~Kurkela, M.~Strickland and A.~Vuorinen,
Phys. Rept. \textbf{880}, 1 (2020).

\bibitem{Rothkopf:2019ipj}
A.~Rothkopf,
Phys. Rept. \textbf{858}, 1 (2020).

\bibitem{Braaten:1991gm}
E.~Braaten and R.~D.~Pisarski,
Phys. Rev. D \textbf{45}, R1827 (1992).

\bibitem{Caswell:1985ui}
W.~E.~Caswell and G.~P.~Lepage,
Phys. Lett. B \textbf{167}, 437 (1986).

\bibitem{Pineda:1997bj}
A.~Pineda and J.~Soto,
Nucl. Phys. B Proc. Suppl. \textbf{64}, 428 (1998).

\bibitem{Brambilla:2008cx}
N.~Brambilla, J.~Ghiglieri, A.~Vairo and P.~Petreczky,
Phys. Rev. D \textbf{78}, 014017 (2008).

\bibitem{Brambilla:2004jw}
N.~Brambilla, A.~Pineda, J.~Soto and A.~Vairo,
Rev. Mod. Phys. \textbf{77}, 1423 (2005).

\bibitem{Pineda:2011dg}
A.~Pineda,
Prog. Part. Nucl. Phys. \textbf{67}, 735 (2012).

\bibitem{Rebhan:1993az}
A.~K.~Rebhan,
Phys. Rev. D \textbf{48}, R3967 (1993).

\bibitem{Shi:2015tmz}
C.~Y.~Shi, J.~Q.~Zhu, Z.~L.~Ma and Y.~D.~Li,
Chin. Phys. Lett. \textbf{32}, 121201 (2015).

\bibitem{Zhu:2015edf}
J.~Q.~Zhu, Z.~L.~Ma, C.~Y.~Shi and Y.~D.~Li,
Nucl. Phys. A \textbf{942}, 54 (2015).



\bibitem{Manuel:2016wqs}
C.~Manuel, J.~Soto and S.~Stetina,
Phys. Rev. D \textbf{94}, 025017,  (2016).

\bibitem{Carignano:2017ovz}
S.~Carignano, C.~Manuel and J.~Soto,
Phys. Lett. B \textbf{780}, 308 (2018).

\bibitem{Leutwyler:1980tn}
H.~Leutwyler,
Phys. Lett. B \textbf{98}, 447 (1981).

\bibitem{Brambilla:1999xj}
N.~Brambilla, A.~Pineda, J.~Soto and A.~Vairo,
Phys. Lett. B \textbf{470}, 215 (1999).

\bibitem{Kiyo:2013aea}
Y.~Kiyo and Y.~Sumino,
Phys. Lett. B \textbf{730}, 76 (2014).

\bibitem{Anzai:2018eua}
C.~Anzai, D.~Moreno and A.~Pineda,
Phys. Rev. D \textbf{98}, 114034 (2018).

\bibitem{Larsen:2019zqv}
R.~Larsen, S.~Meinel, S.~Mukherjee and P.~Petreczky,
Phys. Lett. B \textbf{800}, 135119 (2020).

\bibitem{Carrington:2025cnv}
M.~E.~Carrington, C.~Manuel and J.~Soto,
PoS \textbf{QNP2024}, 109 (2025).

\bibitem{Bazavov:2012ka}
A.~Bazavov, N.~Brambilla, X.~Garcia Tormo, i, P.~Petreczky, J.~Soto and A.~Vairo,
Phys. Rev. D \textbf{86}, 114031 (2012).

\bibitem{Lucha:1998xc}
W.~Lucha and F.~F.~Schoberl,
Int. J. Mod. Phys. C \textbf{10}, 607 (1999).

\bibitem{Aarts:2014cda}
G.~Aarts, C.~Allton, T.~Harris, S.~Kim, M.~P.~Lombardo, S.~M.~Ryan and J.~I.~Skullerud,
JHEP \textbf{07}, 097 (2014).



\bibitem{Kaczmarek:1999mm}
O.~Kaczmarek, F.~Karsch, E.~Laermann and M.~Lutgemeier,
Phys. Rev. D \textbf{62}, 034021 (2000).

\bibitem{Kaczmarek:2002mc}
O.~Kaczmarek, F.~Karsch, P.~Petreczky and F.~Zantow,
Phys. Lett. B \textbf{543}, 41 (2002).

\bibitem{Petreczky:2004pz}
P.~Petreczky and K.~Petrov,
Phys. Rev. D \textbf{70}, 054503 (2004).

\bibitem{Kaczmarek:2005ui}
O.~Kaczmarek and F.~Zantow,
Phys. Rev. D \textbf{71}, 114510 (2005).

\bibitem{Maezawa:2007fc}
Y.~Maezawa \textit{et al.} [WHOT-QCD],
Phys. Rev. D \textbf{75}, 074501 (2007).

\bibitem{Burnier:2009bk}
Y.~Burnier, M.~Laine and M.~Vepsalainen,
JHEP \textbf{01}, 054 (2010)
[erratum: JHEP \textbf{01}, 180 (2013)].

\bibitem{Ekstedt:2023anj}
A.~Ekstedt,
JHEP \textbf{06}, 135 (2023).

\bibitem{Gorda:2023zwy}
T.~Gorda, R.~Paatelainen, S.~S\"appi and K.~Sepp\"anen,
JHEP \textbf{08}, 021 (2023).

\bibitem{Titard:1993nn}
S.~Titard and F.~J.~Yndurain,
Phys. Rev. D \textbf{49}, 6007 (1994).


\end{thebibliography}
\end{document}